\documentclass[fleqn,a4paper,12pt]{article}
\usepackage{amssymb}

\usepackage{amsmath}
\usepackage[dvips]{graphicx}
\usepackage{epsfig}

\newcommand{\bab}{\end{gather}}
\newcommand{\ri}{{\mathrm i}}
\newcommand{\p}{\partial}
\newcommand{\bea}{\begin{array}}
\newcommand{\eea}{\end{array}}
\newcommand{\beg}{\begin{gather}}

\catcode`@=11 \long
\def\@caption#1[#2]#3{\par\addcontentsline{\csname
ext@#1\endcsname}{#1} {\protect\numberline{\csname
the#1\endcsname}{\ignorespaces #2}} \begingroup \small
\@parboxrestore \@makecaption{\csname fnum@#1\endcsname}
{\ignorespaces #3}\par \endgroup} \catcode`@=12

\newcommand{\Q}{\mathbb{Q}}

\newcommand{\la}{\label}

\catcode`@=11 \long
\def\@caption#1[#2]#3{\par\addcontentsline{\csname
ext@#1\endcsname}{#1} {\protect\numberline{\csname
the#1\endcsname}{\ignorespaces #2}} \begingroup \small
\@parboxrestore \@makecaption{\csname fnum@#1\endcsname}
{\ignorespaces #3}\par \endgroup} \catcode`@=12

\begin{document}

\allowdisplaybreaks
 \begin{titlepage} \vskip 2cm

\begin{center} {\Large\bf Integrable and superintegrable quantum mechanical systems with position dependent
masses invariant with respect to one parametric Lie groups. 1. Systems with cylindric symmetry.}

 \vskip 3cm {\bf {A. G. Nikitin }\footnote{E-mail:
{\tt nikitin@imath.kiev.ua} }
\vskip 5pt {\sl Institute of Mathematics, National Academy of
Sciences of Ukraine,\\ 3 Tereshchenkivs'ka Street, Kyiv-4, Ukraine,
01024,\\and
Universit´a del Piemonte Orientale,\\
Dipartimento di Scienze e Innovazione Tecnologica,\\
viale T. Michel 11, 15121 Alessandria, Italy\\}}\end{center}
\vskip .5cm \rm
\begin{abstract} { Cylindrically symmetric quantum mechanical
systems with position dependent masses (PDM) admitting at least one second order integral of motion are classified. It is proved that there exist 68 such systems which are inequivalent. Among them there are twenty seven   superintegrable and twelve maximally superintegrable. The arbitrary elements of the correspondinding Hamiltonians (i.e.,masses and potentials) are presented explicitly. }
\end{abstract}
\end{titlepage}
\section{Introduction\label{int}}

Integrals of motion belong to  very important part of classical and quantum mechanics. Just the existence of the sufficient number of integrals of motion for a Hamiltonian system makes it integrable or exactly solvable, and we are not supposed to calculate its approximate solutions to describe its behavior. Moreover, sometimes it is possible to describe  the main physical properties of the mentioned system  using its integrals of motion an d ignoring the motion equations. A classical example of such situation was presented by Pauli who had calculated the energy spectrum of the Hydrogen atom using its integrals of motion forming the Laplace-Runge-Lentz vector \cite{Pauli}. Ad it was  done before the discovery of Schr\"odinger equation!

The systematic search for integrals of motion admitted by the Schr\"odinger equation equations started with papers  \cite{wint1}, \cite{BM} were all inequivalent  second order  integrals of motion for the 2d one particle quantum systems had been classified. And it needed 24 years to extend this result to the 3d case \cite{ev1}, \cite{ev2}.

Papers  \cite{wint1} and \cite{BM} were indeed seminal. They were followed by a great number of research works, see, e.g. survey \cite{wint01}. In particular, integrable and superintegrable system with matrix potentials have been classified for both spin-orbit \cite{w6, w8} and Pauli type interactions \cite{N1, N5}.   One of the modern trends is to search for the third and even arbitrary order integrals of motion \cite{wintL}, see also  \cite{AGN1} where the determining equations for such integrals were presented.

The higher  (in particular, second) order integrals of motion  are requested
for  description of systems admitting solutions in separated
variables \cite{Miller} just such integrals of motion  characterize integrable and superintegrable systems
\cite{wint01}. Let us mention also the nice conjecture of   Marquette and Winternitz
\cite{Mar}
which predicts a surprising connection of higher order
superintegrability  with the soliton theory.

In any case the integrable and superintegrable systems of the standard quantum mechanics belong to the well developed research field which, however, still have some white spots. Among them is the classification of arbitrary order symmetry operators of generic form, which still are described only at the level of determining equations \cite{AGN1}.

One more important research field which is closely related to the mentioned one is the classification of second order
integrals of motion for quantum mechanical systems with position dependent mass. Such systems
 are used in many branches of modern theoretical physics, whose list can be found,
e.g., in   \cite{Rosas, Roz}. Their symmetries and integrals of motion were studied much less than those ones for the standard SE. However, the classification of Lie symmetries of the PDM Schr\"odinger equations with scalar potentials have been obtained already \cite{NZ, NZ2, AGN}.

Second  order integrals of motion for 2d  PDM SEs are perfectly classified  \cite{Kal1, Kal4}. The majority of such systems admits also  at least one continuous  Lie symmetry. The two dimensional
second-order (maximally) superintegrable systems for Euclidean 2-space had been classified even algebraic geometrically \cite{Kress}.

The situation with the 3d systems is still  indefinite.  At the best of my knowledge the completed classification results were presented only for the maximally superintegrable (i.e., admitting the maximal possible number of integrals of motion) systems \cite{Kal5, Kal31}, and (or) for the system whose integrals of motion are supposed to satisfy some special conditions like the functionally linearly dependence \cite{bern}. The nondegenerate
systems, i.e, those ones which have have 5 linearly independent, contained in 6 linearly independent (but functionally dependent) 2nd order integrals of motion are known \cite{Cap}, see also \cite{Vol} for
the contemporary trends in this field. In addition, a certain progress can be recognized in the
classification of the so called semidegenerate systems which admit five linearly independent integrals of motion and whose potentials are linear combinations of three functionally independent
terms \cite{Mil}.

 Surely, the maximally superintegrable  systems are very important and interesting. In particular, they admit solutions in multi separated coordinates \cite{Ra, Bala2, Rag1}. On the other hand, there are no reasons to ignore the PDM systems which admit second order integrals of motion but are not necessary maximally superintegrable. And just such systems are the subject of our study.

In view of the complexity of the total classification of  integrals of motion for 3d PDM systems   it is reasonable to separate this generic problem to well defined subproblems which can have their own  values. The set of such  subproblems can be treated as optimal one if solving them step by step we can obtain the complete classification of PDM systems admitting integrals of motion.
We choose the optimal set of subproblems  in the following way.

The first subproblem  consists in the  classification of  the PDM systems  admitting  the first order integrals of motion. This problem is already  solved, refer to \cite{NZ}.

The first order integrals of motion are nothing but generators of Lie symmetries. The important aspect of the results presented in  \cite{NZ} is the complete description of possible
Lie symmetry groups which can be admitted by the stationary PDM Schr\"odinger equation.
And this property, i.e., the presence of Lie symmetry, can be effectively used to separate the the problem
of the classification of the PDM systems admitting second order integrals of motion for PDM
systems to a well defined subproblems corresponding to the fixed symmetries.

It was shown in \cite{NZ} the PDM Schr¨odinger equation can admit six, four, three, two or
one parametric Lie symmetry groups. In addition, there are also such equations which have no Lie
symmetry. In other words, there are six well defined classes of such equations which admit n-parametric Lie groups with $n=6, 4, 3, 2, 1$ or do not have any Lie symmetry. And it is a natural idea to search for second order integrals of
motion consequently for all these classes.

The systems admitting six- or four-parametric Lie groups are not too interesting since the related Hamiltonians cannot include non-trivial potentials. That is why we started our research with the case of three-parametric groups. The classification of the corresponding PDM systems admitting second order integrals of motion was obtained in  \cite{AG}. There were specified 38 inequivalent PDM systems together with their integrals of motion. The majority of them are new systems which are not not maximally superintagrable.

Notice that the superintegrable 3d  PDM systems invariant with respect to the 3d rotations
have been classified a bit earlier in paper \cite{154} where their supersymmetric aspects were discussed also. For relativistic aspects of superintegrability see \cite{AG2}, \cite{AG3}.

The systems admitting two-parametric Lie groups and second order integrals of motion have been classified in \cite{AG1}. We again find a number of new systems in addition to the known maximally superintegrable ones.

The natural next step is to classify the systems which admit one parametric Lie groups and second order integrals of motion. As it is shown in \cite{NZ}, up to equivalence these groups are reduced to dilatations, shifts along the fixed coordinate axis, rotations around this axis and some specific combinations of the mentioned transformations. We will conventionally   call them the natural and combined symmetries respectively.

In the present paper we start the systematic search for integrable and superintegrable   PDM systems admitting one parametric Lie symmetry groups.  Namely, we present the classification of the mentioned systems which are invariant w.r.t. the rotations around the fixed coordinate axis. The number of such systems appears to be rather extended, and to keep the reasonable size of paper we restrict ourselves to this particular symmetry, while the systems with the other  Lie symmetries will be presented in the following papers. Notice that the number of inequivalent systems which admit the other one parametric  symmetry groups is much less extended than in the case of the cylindric symmetry.

In spite of the fact that the usual strategy   in studying of superintegrable systems with PDM is to start with the classical Hamiltonian systems an then quantize them if necessary, we deal directly with quantum mechanical systems. This way is more complicated but it guaranties obtaining of all integrals of motion including those ones which can disappear in the classical limit \cite{Hit}.

The main result of the present paper consist in the complete classification of integrable, superintegrable and maximally superintegrable PDM systems with cylindric symmetry. In addition, we optimise the algorithm of solution of the related determining equations which can be used for a classification of other PDM systems.

\section{Formulation of the problem}
We are studying the  stationary  Schr\"odinger equations with position dependent mass of
the following generic form:
\begin{gather}\la{se}
    H \psi=E \psi,
\end{gather}
where
\begin{gather}\la{A1}  H=\frac14(m^\alpha p_a m^{\beta}p_am^\gamma+
m^\gamma p_a m^{\beta}p_am^\alpha)+ \hat V(x)\end{gather}
where $p_a=-i\p_a$, $m=m({\bf x})$ is a function of spatial variables ${\bf x}=(x_1,x_2,x_3)$, associated with the position dependent mass, $\alpha, \beta$  and $\gamma$ are the so called
ambiguity parameters satisfying the condition $\alpha+\beta+\gamma=-1$, and and summation from 1 to 3 is
imposed over the repeating index $a$.

There are various physical speculations how to fix the ambiguity parameters in the particular models based on equation (\ref{A1}) . However, the systems with different values of these parameters are mathematically equivalent up to redefinition of potentials $V({\bf x})$. To simplify the following calculations we will fix them in the following manner: $\beta=0, \gamma=\alpha=-\frac12$ and denote $f=f({\bf  x})=\frac1{2m({\bf x})}$. As a result hamiltonian (\ref{A1}) is reduced to the following form
 \begin{gather}\la{H} H=f^\frac12p_ap_af^\frac12+ V({\bf x}).\end{gather}

In addition we consider the version $\alpha=\gamma=0, \beta=-1$ which corresponds to the following form of hamiltonian (\ref{A1}):
\begin{gather}\la{Ha} H=p_afp_a+ \hat V.\end{gather}

Operators (\ref{Ha}) and (\ref{H}) are equal one to another provided potentials $V=V({\bf x})$ and $\hat V=\hat V({\bf x})$ satisfy the following condition:
\begin{gather}\la{Hb}V=\hat V+V^k\end{gather}
where
\begin{gather}\la{Hc}V^{(k)}=\frac1{4f}((\p_1 f)^2+(\p_2 f)^2+
(\p_3 f)^2)
-\frac12\Delta f\end{gather}
and $\Delta$ is the Laplace operator.

Formula (\ref{Hc}) represents an example of {\it kinematical} potentials which can be reduced to zero by the rearranging  the   ambiguity parameters.

The particular form (\ref{H}) of the hamiltonian is convenient for study of its symmetries and integrals of motion. Moreover, more generic formulations including the arbitrary ambiguity parameters (refer, e.g. to \cite{Roz}) are mathematically equivalent to (\ref{H}).

In paper \cite{NZ} all equations (\ref{se}) admitting at least one first order integral of motion was found.
 Such integrals of motion generate  Lie groups which leave the
equations invariant.  In accordance with  \cite{NZ} there are six inequivalent Lie symmetry groups which can
 be accepted by the PDM Schr\"odinger equations. They include three "natural" groups,  rotation around the third coordinate axis, shift along this axis
 and dilatation groups. In addition, we can fix three combined symmetries which are superpositions
 of rotations and shifts, rotations and dilatations, and shifts, rotations and conformal transformations.

The generic form  of the corresponding inverse masses $f$ and potentials $V$ can be represented by  the following formulae \cite{NZ}:
\begin{gather}\la{f_V2}f=F(\tilde r,x_3), \quad V=G(\tilde r,x_3),\\\la{f_V1}f=F(x_1,x_2), \quad V=V(x_1,x_2),\\
\la{f_V3}f= r^2F(\varphi,\theta), \quad V=V(\varphi,\theta)\end{gather}
where $F(.)$ and $V(.)$ are arbitrary functions whose arguments are fixed in the brackets,
\begin{gather*} r=(x^2_1+x_2^2+x_3^2)^\frac12,\quad  \tilde r=(x^2_1+x_2^2)^\frac12,\quad \varphi=\arctan\left(\frac{x_2}{x_1}\right), \ \ \theta=\arctan\left(\frac{\tilde r}{x_3}\right).\end{gather*}

In the present  paper we classify the  PDM systems which admit one of the mentioned natural symmetries,
namely the rotations around the third coordinate axis,  and, in addition, have at least one second order integral
of motion. The generic form  of the corresponding inverse masses $f$ and potentials $V$ are represented in
(\ref{f_V2}).

Equations (\ref{se}), (\ref{H}) with arbitrary parameters presented in (\ref{f_V2}) admit the following first order integrals of motion
\begin{gather}\la{IM1}L_3=x_1p_2-x_2p_1\end{gather}
which is nothing but the third component of the orbital momentum.

Our goal is to fix such systems (\ref{se}), (\ref{f_V2}) which, in addition to their Lie symmetries, admit second order integrals of motion whose generic form is:
\begin{equation}\label{Q}
    Q=\p_a\mu^{ab}\p_b+\eta
\end{equation}
where $\mu^{ab}=\mu^{ba}$ and $\eta$ are unknown functions  of $\bf
x$ and summation from 1 to 3 is imposed over all repeating indices.

Operators (\ref{Q}) are formally hermitian. In addition, just representation (\ref{Q}) leads to the most compact and simple systems of determining equations for unknown parameters   $\mu^{ab}$ and $\eta$.

By definition, operators  $Q$ should commute with $H$:
\begin{equation}\label{HQ}[ H,Q]\equiv  H Q-Q H=0.\end{equation}
Evaluating the commutator in (\ref{HQ})
and equating to zero the coefficients for the linearly independent differential operators $\p_a\p_b\p_c$ and $\p_a$ we come to the following determining equations
\begin{gather}\la{m0}5\left(\mu^{ab}_c+\mu^{ac}_b+ \mu^{bc}_a\right)=
\delta^{ab}\left(\mu^{nn}_c+2\mu^{cn}_n\right)+
\delta^{bc}\left(\mu^{nn}_a+2\mu^{an}_n\right)+\delta^{ac}
\left(\mu^{nn}_b+2\mu^{bn}_n\right),\\
\la{m1}
  \left(\mu^{nn}_a+2\mu^{na}_n\right)f-
5\mu^{an}f_n=0,\\\la{m2}\mu^{ab}V_b-f\eta_a+F^a=0\end{gather}
  where $\delta^{bc}$ is the Kronecker delta, $f_n=\frac{\p f}{\p x_n}, \ \mu^{an}_n=\frac{\p \mu^{an}}{\p x_n }$, etc., and summation is imposed over the repeating indices $n$ over the values $n=1,2,3$.
Moreover, 
\begin{gather}\la{e20b}F^a=(f\mu^{ak}_{km}-\mu^{mn}f_{na})_m\end{gather}
  for the  Hamiltonian of form (\ref{Ha}) and
\begin{gather}\la{e20c}F^a=(f\mu^{ak}_{km}-\mu^{mn}f_{na})_m+\mu^{ab}V^{(k)}_b\end{gather}
for the  Hamiltonian of form (\ref{Hb}).

 Equations (\ref{m0}) , (\ref{m1}) and (\ref{m2})  give the necessary and sufficient conditions for commutativity of operators $H$ (\ref{H}) and (\ref{Ha}) with $Q$ (\ref{Q}) \cite{AG1}.

\section{Evolution of the determining equations}

A particular solution of equations (\ref{m0}) is $\mu^{ab}=\mu_0^{ab}$ where
\begin{gather}\la{K0}\mu_0^{ab}=\delta^{ab}g({\bf r})\end{gather}
 with arbitrary function $g({\bf r})$.

Whenever tensor $\mu_0^{ab}$ is nontrivial, the determining equations (\ref{m1})
represent the coupled system of three {\it nonlinear} partial differential equation equations for
two unknowns $g(x)$ and $f({\bf x})$. Fortunately, this system can be linearizing by introduction of the
new dependent variables
\begin{gather}\la{MN}M=\frac1f, \ N=\frac{g}f\end{gather}
 which reduces (\ref{m1})   to the following form:
\begin{gather}\la{me1}
 \left(\mu^{nn}_a+2\mu^{na}_n\right)M+
5(\mu^{an}M_n+N_a)=0.\end{gather}

We see that for variables (\ref{MN}) the determining equation (\ref{m1}) is linear. Equation (\ref{m2}) in its turn
 can be effectively linearised by introducing the following new dependent variables $\tilde M$ and $R$:
 \begin{gather}\la{MN!}\tilde M=MV,\ R=\frac{N\tilde M}M-\eta\end{gather}
 which reduce (\ref{m2}) to the following equation:
 \begin{gather}\la{me2}
 \left(\mu^{nn}_a+2\mu^{na}_n\right)\tilde M+
5(\mu^{an}\tilde M_n+R_a)=0\end{gather}
which simple coincides with (\ref{me1}). Surely, it does not mean that $M$ and $N$ coincide with $\tilde M$ and
$\tilde N$ respectively, since these functions can include different arbitrary elements, say, different integration constants. In accordance with (\ref{MN}) and  (\ref{MN!}) the related inverse mass potential have the form:
\begin{gather}\la{Pot}
f=\frac 1{M}, \ V=\frac{ \tilde M}{M}\end{gather}
where $\tilde M$ and $M$ are different solutions of the same equation, while the corresponding functions $g$ and $\eta$ are expressed via $M$ and $\tilde M$ in the following manner:
\begin{gather}\la{MN!!} g= \frac{N}M, \ \eta=-\frac{N\tilde M}M-
 R.\end{gather}

We see that to find the admissible inverse mass and potential it is sufficient to solve the only linear equation
(\ref{me1}) and then find the desired functions $f, V, g $ and $\eta$ using definitions (\ref{Pot})
 and (\ref{MN!!}).

Just linearised determining equation (\ref{MN}) together with the mentioned definitions will be used in the
following to solve our classification problem.

Let us represent generic  integral of motion  (\ref{Q}) in terms of new dependent variables $M, N, \tilde M$ and $\tilde N$ (refer to (\ref{MN}) and (\ref{MN!}))
\begin{gather}\la{NV}Q=P_a\mu^{ab}P_b+(N\cdot H)-R\end{gather}
where we denote
\begin{gather}\la{N*H}(N\cdot H)=P_a(Nf)P_a+NV.\end{gather}

The latter definition includes  a Hermitized product of function $N$ with Hamiltonian  (\ref{H}).

It is necessary to note than whenever function $g({\bf r})$ is equal to zero, i.e.,
tensor  $\mu_0^{ab}$ is trivial, the above presented speculations are forbidden. We still can deal with the determining equations (\ref{me1}), but  we are supposed to
deal with the initial determining equations  (\ref{m2}) instead of (\ref{me2}).

 \section{Equivalence relations}

 An  important  step of our classification problem  is the definition of equivalence relations which will be presented in this section.

    Non degenerated changes of dependent and independent variables of  equations  (\ref{se}), (\ref{H}) are called  equivalence transformations provided they keep their  generic form up to the explicit expressions for the arbitrary elements $f$  and $V$.  They have the structure of a continuous group which however can be extended by  some discrete elements.
    Let us remind that a particular subset of the equivalence transformations are invariance transformation which by definition keep the mentioned arbitrary elements uncharged.

It was shown   in \cite{NZ} that the maximal continuous equivalence group of equation  (\ref{se}) is  the group of conformal transformations of the 3d Euclidean space which we denote as C(3). The corresponding Lie algebra is a linear span of the following first order differential operators \cite{NZ}:
\begin{gather}\label{QQ}\begin{split}&
 P^{a}=p^{a}=-i\frac{\partial}{\partial x_{a}},\quad L^{a}=\varepsilon^{abc}x^bp^c, \\&
D=x_n p^n-\frac{3\ri}2,\quad K^{a}=r^2 p^a -2x^aD,\end{split}
\end{gather}
where $r^2=x_1^2+x_2^2+x_3^2$  and $p_a=-i\frac{\p}{\p x_a}.$ Operators $ P^{a},$ $ L^{a},$ $D$ and $ K^{a}$ generate shifts, rotations, dilatations and pure conformal transformations respectively.

In addition, equation (\ref{se}) is form invariant with respect to the following discrete transformations:
 \begin{gather}\la{IT} x_a\to
\tilde x_a=\frac{x_a}{r^2},\quad \psi({\bf x})\to \tilde x^3\psi(\tilde{\bf x})\end{gather}
where $\tilde x=\sqrt{\tilde x_1^2+\tilde x_2^2+\tilde x_3^2}.$

Notice that the related Lie algebra c(3) is isomorphic to the algebra so(1,4) whose basic elements $S_{\mu\nu}$ can be expressed via generators  (\ref{QQ}) as:
 \begin{gather}\la{so} S_{ab}=\varepsilon_{abc}L_c, \quad S_{4a}=\frac12(K_a-P_a),\quad S_{0a}=\frac12(K_a+P_a), \quad S_{04}=D\end{gather}
 where $a, b=1, 2, 3.$
 The related Lie group SO(1,4) is the Lorentz group in (1+4)-dimensional space. The discrete transformation (\ref{IT})  anticommutes with $S_{4a}$ and $S_{40}$  but commutes  with the remaining generators (\ref{so}). Thus its action on operators (\ref{so}) can be represented as follows:
 \begin{gather}\la{soo}S_{4a}\to - S_{4a},\ S_{04}\to - S_{04}, \ S_{0a}\to  S_{0a}, \ S_{ab} \to S_{ab}.\end{gather}

The presented speculations are valid for an abstract system  (\ref{se}) which is free of any additional constrains.
However, for the systems whose arbitrary elements satisfy condition (\ref{f_V2})  the equivalence group is
reduced since it is supposed that it does not change the invariance groups of these equations.
 It means that the set of  generators  (\ref{so}) should be reduced to such ones which  commute
 with $L_3$. There are four the generators satisfying this condition, namely:

 \begin{gather}\la{N3}  \ P_3,\ L_3,\ K_3,\ D.
 \end{gather}
 They generate the reduced  equivalence algebra so(2,1)$\oplus$e(1) where e(1) includes the only basis
 element $L_3$.

Thus, in comparison with the generic case, for special arbitrary elements presented in (\ref{f_V2}) the admissible
continuous equivalence transformations are reduced to group  SO(2,1) extended by rotations w.r.t. the
third coordinate axis.
  However, the admissible discrete transformations are extended by
 the reflection of one out of two  the independent
 variables $x_1$ and $x_2$, i.e.,
 \begin{gather}\la{so1} x_1\to -x_1, \ x_2\to x_2, \ x_3\to x_3 \end{gather}
 or
 \begin{gather}\la{so2}\ x_1\to x_1, \ x_2\to -x_2,\  x_3\to x_3 \end{gather}
 which  keep the related equation (\ref{se}) invariant. These discrete transformations can be added to
 the universal discrete transformation (\ref{IT}).

 The presented equivalence relations will be used in the following to simplify calculations and and to optimize the representation of the classification results.

\section{Identities in the extended enveloping algebra of c(3)}

It was noted in \cite{AG1} that integrals of motion (\ref{Q}) where $\mu^{ab}$ Killing tensors, i.e., solutions of equations   can be represented as bilinear combinations of the basic elements of algebra c(3) (\ref{QQ}) added by the special term with $\mu^{ab}=\delta_{ab}g({\bf x})$
and potential term $\eta$. In other words they admit the following representation:
\begin{gather}\la{QQQ}Q=c^{\mu \nu,\lambda\sigma}\{S_{\mu \nu},S_{\lambda \sigma}\}+P_a g({\bf x})P_a+\eta\end{gather}
where $S_{\mu \nu}$ are generators (\ref{so}) and $c^{\mu \nu,\lambda\sigma}$ are numeric parameters.

In accordance with (\ref{QQQ}) the second order integrals of motion of the PDM Schr\"odinger equation belong to the enveloping algebra of their equivalence algebra, i.e., c(3), extended by special terms $p_a g({\bf x})p_a$. Notice that the same is true for the any order integrals of motion with the appropriate generalization of the extending term.

 Representation (\ref{QQQ}) is very important. Being combined with the equivalence relations discussed in the previous section it enables essentially simplify both the calculations  and the representation of their results. In addition to the equivalence relations we will use numerous identities in the enveloping algebra of algebra c(3) which take place for its particular  realization (\ref{QQ}). These identities are presented in the following formulae:
 \begin{gather}\la{Id2}\{P_a,D\}+
\varepsilon_{abc}\{P_b,L_c\}=2P_cx_aP_c,\\\la{Id0}L_1^2+L_2^2+L_3^2+D^2
=P_ar^2P_a, \\\la{Id3}
  \{L_a,L_b\}+\{P_a,K_b\}-\delta^{ab}(L_1^2+L_2^2+L_3^2)=2Q^{ab}\\
\begin{split}&P_1^2+P_2^2+P_3^2=P_aP_a,\\
&\{P_a,K_b\}-\{P_b,K_a\}=2\varepsilon_{abc}L_cD,\\&P_1L_1+P_2L_2+P_3L_3=0
\end{split}\la{Id1}\end{gather}
where $Q^{ab}=-P_cx_ax_bP_c.$

The message given by relations  (\ref{Id2})-(\ref{Id1}) is that the terms in the l.h.s. can be treated
as linearly dependent whenever they are included into second order integral of motion (\ref{QQQ})
since the latter one includes a yet indefinite term of the kind presented in the r.h.s. of equations
 (\ref{Id2})-(\ref{Id1}).

We will use relations   (\ref{Id2})-(\ref{Id1}) to produce maximally  compact presentations for the integrals of motion.

\section{Solution of determining equations}

The autonomous subsystem (\ref{m0}) defines the conformal Killing tensor which is the fourth order
polynomial in variables $x_a$ and includes an arbitrary function which multiplies the Kronecker delta. The
explicit expression for this polynomial are presented in (\ref{K0}) and in the following formulae (see, e.g., \cite{Kil}) :
\begin{gather}\la{K1}
\mu^{ab}_1=\lambda_1^{ab},\\
\la{K2}
\mu^{ab}_2=\lambda_2^a x^b+\lambda_2^b x^a-2\delta^{ab}\lambda_2^c
x^c,
\\\la{K3}
\mu^{ab}_3=(\varepsilon^{acd}\lambda_3^{cb}+ \varepsilon^{bcd}
\lambda_3^{ca})x^d,\\
\la{K4}
\mu^{ab}_4=(x^a\varepsilon^{bcd}+x^b\varepsilon^{acd}) x^c\lambda^d_4,
\\\la{K5}\mu^{ab}_5=\delta^{ab}r^2+
k (x^ax^b-\delta^{ab}r^2),\\\la{K6}
\mu^{ab}_6=\lambda_6^{ab}r^2-(x^a\lambda_6^{bc}+x^b\lambda_5^{ac})x^c-
\delta^{ab}\lambda_6^{cd}x^cx^d,\\
\la{K7}\mu^{ab}_7=(x^a\lambda_7^b+x^b\lambda_7^a)r^2-4x^ax^b\lambda_7^c x^c+
\delta^{ab}
 \lambda_7^c x^cr^2,
\\\la{K8}\mu^{ab}_8= 2(x^a\varepsilon^{bcd} +x^b\varepsilon^{acd})
\lambda_8^{dn}x^cx^n- (\varepsilon^{ack}\lambda_8^{bk}+
\varepsilon^{bck}\lambda_8^{ak})x^cr^2,\\
\mu^{ab}_9=\lambda_9^{ab}r^4-2(x^a\lambda_9^{bc}+x^b\lambda_9^{ac})x^cr^2+
(4x^ax^b+\delta^{ab}r^2)\lambda_{9}^{cd}x^cx^d+
\delta^{ab}\lambda_{9}^{cd}x^cx^dr^2\la{K9}
\end{gather}
where $r=\sqrt{x_1^2+x_2^2+x_3^2}$, $\lambda_n^a $ and $\lambda_n^{ab}$  are
arbitrary parameters, satisfying the conditions $\lambda_n^{ab}=\lambda_n^{ba}, \lambda_n^{bb}=0$.

Thus we have to search for solutions of the determining equations  (\ref{m1})  where $\mu^{ab}$
are linear combinations of tensors (\ref{K0}), (\ref{K1})-(\ref{K9}).

Formulae (\ref{K0}) and (\ref{K1})-(\ref{K9})  include an arbitrary function $g({\bf x})$ and
  35 arbitrary parameters $\lambda_n^a$ and $\lambda_n^{ab}$, $a, b =1,2,...,9$. In addition, we have eight
 coefficients which appear in an arbitrary linear combination of tensors (\ref{K0}),  and  three more unknown
function, i.e., $f$, $V$ and $ \eta$. Thus the problem of the complete classification  of the 3d PDM systems
admitting second order integrals of motion looks to be huge.  However, our strategy is to solve it step by step
for the systems admitting three, two and one parametric Lie groups, and, finally, for the systems which do
not admit any Lie symmetry. The first two steps have been already done in papers  \cite{AG} and \cite{AG1}.
 The third step is the subject of the current  paper.

We are studying the  PDM systems which are invariant with respect to rotations around the fixed axis (say, the third one). In accordance with  (\ref{f_V2}) the corresponding Hamiltonian (\ref{H}) is reduced to the following form:
\begin{gather}\la{H1}H=p_af(\tilde r,x_3)p_a+V(\tilde r, x_3) \end{gather}
where $\tilde r=\sqrt{x_1^2+x_2^2}$.

As it was noted in Section 3 the equivalence group of equation (\ref{se}) is reduced to  SO(2,1)$\otimes$E(1)
 provided the related Hamiltonian (\ref{H}) has the reduced form (\ref{H1}).
 The corresponding infinitesimal operators are presented in (\ref{N3}). In addition, there is the discrete
 equivalence transformations (\ref{IT}), (\ref{so1}) and (\ref{so2}).

However, the symmetry fixed above makes it possible to decouple the second order integrals of motion
to the following three subclasses: scalars, vectors and second rank tensors with respect to rotations around the third coordinate axis. We will consider them consequently.

\subsection{Scalar integrals of motion}

The scalar integrals of motion are generated by the Killing tensors (\ref{K1})-(\ref{K9}) with the following nontrivial parameters  $\lambda^a_n $ and $\lambda_m^{ab}$:
\begin{gather}\la{Lambdaab}\lambda_n^3\neq0, \ n=2, 4, 5, 7; \ \lambda_m^{33}\neq0, \ m= 1, 3, 6,
8, 9\end{gather}
while the remaining parameters should be identically zeros. These scalars can be expressed via bilinear combinations of generators (\ref{QQ}) in the following manner:
\begin{gather}\la{s03}S_3= P_3 L_3+...,  \ S_4=D L_3+..., S_8=\{L_3,K_3\}+...,\\ \begin{split}&  \la{s04}S_1=P_3^2+..., S_2=\{P_3,D\}+..., S_6=\{P_3,K_3\}+...,\\ &S_5=D^2+..., \ S_7:=\{K_3,D\}+..., \ S_9:=K_3^2+... \end{split} \end{gather}
where the dots denote the differential operators $S_0=P_ag_n(\tilde r,x_3)P_a$  generated by $\mu_0$ (\ref{K0}) and functions $\eta_n$ present in definition (\ref{Q}), and the numbers presented as subindices coincide with the number of the Killing tensor in (\ref{K1}-(\ref{K9}) which generate the scalar.

All operators (\ref{s03}) and (\ref{s04}) are invariant with respect to rotations around the third coordinate axis. However, they have different transformation properties with respect to reflections  (\ref{so1}) and (\ref{so2}). Namely, operators (\ref{s03}) change their sign and so are pseudo scalars while operators (\ref{s04}) remain unchanged and so are true or proper scalars.

The generic scalar integral of motion  is a linear combinations of operators (\ref{s03}) and (\ref{s04}). Surely, such a linear combination cannot include both true and pseudo scalars, and we have one more separation of our integrals to two subclasses.

Let us specify the inequivalent linear combinations of pseudo scalars  (\ref{s03}). First we note that all of them are products of operator $L_3$ and linear combinations of the following operators:
\begin{gather}S_{04}=D, \ S_{03}=\frac12(K_3+P_3), \ S_{43}=\frac12(K_3-P_3)\la{s06}\end{gather}
which satisfy the following relations
\begin{gather}\la{007}[S_{04},S_{03}]=-\ri S_{43},\ [S_{04},S_{43}]=-\ri S_{03},\  [S_{04},S_{43}]=-\ri S_{04}\end{gather} and so form a basis of Lie algebra so(1,2). Moreover, operators (\ref{s06}) commute with $L_3$. It means that the number of inequivalent pseudoscalars which are nothing but the products of linear combinations of operators (\ref{s06}) with operator $L_3$ is equal to the number of inequivalent subalgebras of algebra so(1,2), since just these subalgebras generate our integrals of motion.

The inequivalent subalgebras of algebra $so(1,2)$ are one dimensional and include the following basis elements:
\begin{gather}\la{s061}S_{03}, \ S_{43}, \ S_{03} \pm S_{43}.\end{gather}
Moreover, up to discrete transformation (\ref{IT}) which changes the sign of $S_{03}$ we can restrict ourselves to the positive sign in the last term in (\ref{s061}).
As a result we obtain the following
inequivalent symmetries (\ref{QQQ}) which we present together with the related Killing tensors (\ref{K1})-(\ref{K9}):
\begin{gather}\la{s05} \hat Q_1= P_3L_3+P_a\tilde q_1(\tilde r,x_3)P_a+\eta_1,\ \mu^{ab} = \mu_3^{ab}\end{gather}
and
\begin{gather}\la{s051}\hat Q_2= (K_3+P_3)L_3+P_a\tilde q_2(\tilde r,x_3)P_a+\eta_2,\ \mu^{ab} = \mu_3^{ab}+ \mu_8^{ab},\\
\la{s052}\hat Q_3= (K_3- P_3)L_3+P_a\tilde q_3(\tilde r,x_3)P_a+\eta_3,\ \mu^{ab} = \mu_3^{ab}- \mu_8^{ab}\end{gather}
 we remind that the only nonzero parameters in tensors $\mu_3^{ab}$ and $\mu_8^{ab}$ are $\lambda^{33}_3$ and $\lambda^{33}_8$.

 Let us specify one more pseudo scalar operator
 \begin{gather}\la{s054}\hat Q_4=\{D,L_3\}+P_a\tilde q_4(\tilde r,x_3)P_a, \ \mu^{ab} = \mu_4^{ab}, \lambda_4^1\neq0.\end{gather}
 Operator (\ref{s054}) is equivalent to (\ref{s051}) and can be ignored in the analysis of integrable PDM systems. However, we cannot ignore it in the case of superintegrable systems when it can appear in a combinations with other symmetries.

The specific arguments of functions $q(.)$ in (\ref{s012}) are caused by the requested symmetry of (\ref{s012}) with respect to rotations around the third coordinate axis.

The next task is to specify all inequivalent proper scalars. In accordance with (\ref{s03} ) and (\ref{007}) they belong to the enveloping algebra of algebra so(1,2).
Moreover, it follows from  the first of equations (\ref{Id0}) that the related  Casimir operator of algebra so(1,2) takes the following form:
\begin{gather}\la{s07} S_{03}^2+S_{04}^2-S_{43}^2=P_a\tilde r^2P_a-L_3^2.\end{gather}

Thus the considered integrals of motion are linear combination of basic elements of the mentioned enveloping algebra:
\begin{gather}\la{s08} Q=\sum c_\alpha Q_\alpha\end{gather}
where $\alpha=1,...,6$,
\begin{gather}\la{s09}\begin{split}& Q_1=(S_{03}^2-S_{43}^2),\ Q_2=(S_{03}^2+S_{43}^2),\ Q_3=\{S_{03},S_{43}\},\\&
Q_4=\{S_{03},S_{04}\},\ Q_5=\{S_{04},S_{43}\}, \ Q_6=S_{04}^2-S_{03}^2\end{split}\end{gather}
and $c_1,...,c_6$ are real constants.

Notice that relation (\ref{s07}) can be rewritten in terms of operators $Q_1, Q_2$ and $Q_6$ in the following
way:
\begin{gather}\la{WA}Q_1+3Q_2+2Q_6=2(P_a\tilde r^2P_a-L_3^2).\end{gather}

The l.h.s. of relation  (\ref{WA}) includes the symmetry operator $L_3^2$ which commutes with
Hamiltonian (\ref{H}) by definition and the term $P_a\tilde r^2P_a$ which can be included to the last term of
the generic integral of motion (\ref{QQQ}). It means that operators $Q_1, Q_2$ and $Q_6$  can be treated as
linearly dependent and so one of the coefficients $c_1, c_2$ and $c_3$ can be nullified without loss of
generality. Thus the generic linear combination (\ref{s09}) includes five terms only.

Rather tedious speculations with using the mentioned in the above equivalence relations and identities (\ref{Id2}) (see the preprint \cite{AG5}) for the details) make it possible to specify the following inequivalent scalar  symmetries (\ref{s09}) which we present together with the related Killing tensors:

\begin{gather}\la{S02} Q_1=P_3^2 + P_aq_1(\tilde r,x_3)P_a+\eta^1, \ \mu^{ab}=\mu_1^{ab},\\
\la{S01}Q_2=\{P_3,D\} + P_aq_2(\tilde r,x_3)P_a+\eta^2, \ \mu^{ab}=\mu_2^{ab},\\
\la{S03}
 Q_3=\{P_3,K_3\}+P_aq_3(\tilde r,x_3)P_a, \ \mu^{ab}=\mu_6^{ab}+\eta^3,\\
 Q_4=P_3^2\pm K_3^2+P_aq_4(\tilde r,x_3)P_a+\eta^4,\ \mu^{ab}=\mu_1^{ab}\pm\mu_9^{ab},\la{s015}\\
 \la{s012}\begin{split}&Q_5= (K_3\pm P_3)^2+P_aq_5(\tilde r,x_3)P_a+\eta^5, \ \mu^{ab}=\mu_1^{ab}\pm 2\mu_6^{ab}+\mu_9^{ab},\end{split}\\
 \la{S013}Q_6=\{P_3,K_3\pm P_3\}+P_aq_6(\tilde r,x_3)P_a+\eta^6, \mu^{ab}=\mu_6^{ab}\pm \mu_1^{ab},\\
\la{S015}Q_7=K_3^2\pm P_3^2+ 2n\{K_3,P_3\}+P_aq_7(\tilde r,x_3)P_a+\eta^7, \ \mu^{ab}=\mu^{ab}_9-
\mu^{ab}_1+ 2n\mu^{ab}_6,\\
\la{S016}Q_8=\{D,(K_3\pm P_3)\}+P_aq_8(\tilde r,x_3)P_a+\eta^8, \ \mu^{ab}=\mu_7^{ab}\pm\mu_1^{ab}
\end{gather}
and nonzero parameters in tensors $\mu^{ab}_1, \mu^{ab}_2, \mu^{ab}_6, \mu_7^{ab}$ and $\mu^{ab}_9$ are $\lambda_1^{33},  \lambda_2^{33}, \lambda_6^{33}, \lambda _7^3$ and $\lambda_9^{33}$.

Thus to classify the scalar integrals of motion it is sufficient to solve the determining equations (\ref{m1}) and (\ref{m2}) where $\mu^{ab}$ are the Killing tensors fixed in (\ref{S02}) - (\ref{S016}).

\subsection{Pseudo scalar integrals of motion}

Let us search for solutions of the above defined determining equations for pseudo scalar integrals of motion.

First we present the related determining equations which take   the following form:
\begin{gather}\la{s001}x_2\p_3 M=\p_1 N,\ x_1\p_3 M =-\p_2 N,\ \p_3 N=0 \end{gather}
for $ \hat Q_1,$
\begin{gather}\la{002}\p_3N=0, \ \p_{4}N=0, \\\la{003}\p_\varphi N=\tilde r^2(2x_3\tilde r^2\p_4M+2x_3M)-\frac12(\tilde r^2-x_3^2\pm1)\p_3M
\end{gather}
for $\hat Q_2$ and $\hat Q_3$ were $ \hat Q_1-\hat Q_3$ are operators (\ref{s05})- (\ref{s052}),
 $ \p_3=\frac{\p}{\p_{x_3}},
\p_4=\frac{\p}{\p_{x_4}}, x_4=\tilde r^2$.

Equations (\ref{s001}) are easily integrable and
are solved by the following functions:
\begin{gather}\la{1} M= \frac{\nu x_3+F(\tilde r)}{\tilde r^2},\ N=\frac{\nu}2\varphi \ \text{ for } \hat Q_1,
\end{gather}
where $\varphi=\arctan\left(\frac{x_2}{x_1}\right) $ is the Euler angle, $\nu$ is the integration constant, and $F(\tilde r)$ is an arbitrary function of $\tilde r=\sqrt{x_1^2+x_2^2}.$
The corresponding functions $\tilde M$ and $\tilde N$ which generate  potential $V$ and function $\eta$  are
solutions of the same equation and so have the same generic forms as given in (\ref{1}), but in general with
different parameter $\nu$ and different function $F(\tilde r)$. Substituting the obtained results into
(\ref{MN!!}) we come to functions $f$ and $V$ presented in Item 1 of Table 1.

Equations (\ref{002}) and (\ref{003}) are easily solvable also.
In accordance with (\ref{002}) function $N$ depends on $\varphi$ only. Since $M$ by definition does not
depend on this variable, to solve equation (\ref{003}) $N$ should be a linear function, i.e., $N=c\varphi.$
Then equation (\ref{003}) is reduced to the following one:
\begin{gather}\la{005}2x_3\tilde r^2\p_{4}M+2x_3M-\frac12((\tilde r^2-x_3^2\pm1)\p_3M=c.\end{gather}
 Its solutions are presented
 in Items 2 and 3 of Table 1.

  Notice that these solutions are qualitative different for different sighs before 1 in the   formulae   presented
  above. And these signs are the same as signs for $P_3$ in equations (\ref{s051}) and (\ref{s052}).

Let us fix also the PDM system admitting symmetry $D L_3+...$ which is generated by the Killing tensor
(\ref{K5}) with the only nonzero parameter $\lambda_4^3$.. This symmetry is equivalent to the symmetry
 $L_3(P_3+K_3)+...$  considered above, see equation (\ref{s05}). However, its presentation will be useful in searching for the systems admitting more than one second order integrals of motion. We will not present the calculation details but give the corresponding equations (\ref{me1}):
\begin{gather}\la{NNN}\begin{array}{l}x_4x_3\p_3M+2x_4\p_4M+2M+\p_\varphi N=0,\\\p_3N=0,\ \p_4 N=0, \ \p_{\varphi\varphi}N=0\end{array}\end{gather}
 and their  solutions:
 \begin{gather}\la{006} M=\frac{F(\theta)-2\nu \ln(\tilde r)}{\tilde r^2}\end{gather}
 where $F(\theta)$ is arbitrary function, $\varphi$ and $\theta$  are the Euler angles, $\mu$ and $\nu$ are arbitrary parameters.

 Just these functions together with the corresponding integrals of motion are presented in Item 4 of Table 1.

  Thus we have found all inequivalent  systems with position dependent mass which admit pseudo scalar integrals of motion. They are defined up to pairs of arbitrary functions. The related Hamiltonians commute also with the third component of angular momentum and so the found systems are integrable. We will se that for some particular arbitrary functions these systems are superintegrable.

  \subsection{True  scalar integrals of motion}
  Consider now the integrals of motion which are invariant with respect to the space reflections.  Their  generic form  is given by  formulae (\ref{S02})--(\ref{S015}).

  Let us represent the corresponding determining equations. They  include functions $M$ and $N$ which depend on  $\tilde r$ and $x_3$, but in some cases it is reasonable to treat them as functions of $r$ and $x_3$. To unify the representation we will use the notations $\tilde r^2=x_4$ and $ r^2=x_5$ .

  Substituting the Killing tensors fixed in (\ref{S02})-(\ref{S015}) into (\ref{me1}) we come to tjhe following equations:
\begin{gather}\la{Q1}\text{ For } \ Q_1: \ \p_{\tilde r} N=0,\ \p_3(M+N)=0;\\
\la{Q2} \text { For } \ Q_2: \ \begin{array}{l}\p_3M+2\p_{5}N=0,\\2\p_{ 5}(x_5 M+x_3N)+\p_3(x_3M+N)-2M=0;\end{array}\\
\la{Q3} \text{ For } \ Q_3: \ \begin{array}{l}2\p_{4}(x_3^2M+N)-x_3p_3M=0,\\\p_3(\tilde r^2M+N)-2x_3\tilde r^2\p_4M=0;\end{array}\\
\la{Q4} \text{ For } \ Q_4: \ \begin{array}{l}\p_{4}(N+4x_3^2\tilde r^2M)-W_\pm\p_3 M=0,\\
  \p_3(N+W_\pm^2M)-4x_3\tilde r^2W_\pm\p_{4}M=0, \\W_\pm=\tilde r^2\pm1-x_3^2=x_4-x_3^2\pm1;\end{array}
  \\\la{Q5}\text{ For } \ Q_5: \ \begin{array}{l}\p_{4}(N+4(x_3^2\pm1)\tilde r^2M)-x_3(W_\pm)\p_3M=0,\\\p_3 N+((W\pm)^2+4\tilde r^2)\p_{3} M-4\p_{4}(x_3\tilde r^2 M(W\mp2));\end{array}\\ \la{Q6}\text{ For } \ Q_6:\ \begin{array}{l}(2x_3^2\p_4M-x_3\p_3)M=\p_4N,\\
  (x_4^2-1)(\p_3-2x_3x_4\p_4)M=\p_3N;\end{array}\\
 \la{Q7} \text{ For } \ Q_7:\ \begin{array}{l} x_3(x_4-x_3^2+n)\p_3M+2(N+2-2x_3^2)\p_4(x_4M)=\p_4N,\\
  (x_3^4-2(x_4+n) x_3^2+x_4^2
  -4x_4\pm1)\p_3M\\-4x_3(x_4-x_3^2+n)\p_4x_4M=-\p_3N,\end{array}\\
   \la{Q8} \text{ For } \ Q_8: \ \begin{array}{l} (3x_3^2-x_4\pm1)\p_3M+8x_3\p_4(x_4M)=2\p_4N,\\
   2(x_4-3x_3^2\pm1)(x_4\p_4M+\p_3(x_3M))=-\p_3N\end{array}
 \end{gather}

The systems (\ref{Q1}) and (\ref{Q3}) are easy solvable. They are solved by the following functions:
\begin{gather*}M=F(\tilde r)+G(x_3), \ N=-G(x_3)\end{gather*}
and
\begin{gather*}M=\frac{F(\theta)+G(R)}{r^2}, \ N=-F(\theta)\end{gather*}
where $F(.)$ and $\theta$ are arbitrary functions, $\theta=\arctan\left(\frac{\tilde r}{x_3}\right)$ is the Euler angle.
These solutions generate the inverse masses and potentials  represented in Items 5 and 6 of Table 1

Equations (\ref{Q4}) are a bit more complicated.
Excluding  unknown variable $N$ we obtain the following second order equation for  $M$:
\begin{gather}\la{poly}\p_{33}M-4\p_{55}(x_5M)+2\p_5M=0.\end{gather}
The obtained partial differential equation with variable coefficients appears to be exactly solvable. To discover its exact solutions it is reasonable  to represent them as
\begin{gather}\la{poly1}M=\frac{P}{\sqrt{x_5}} =\frac{P}{r}\end{gather} where $P$ are polynomials in $r$ and $x_3$.

 The only first order polynomial satisfying (\ref{poly}), (\ref{poly}) is $c_1x_3$, the second order polynomial is  $c_2r x_3 +c_3(r^2+x_3^2),$ the third order polynomial is $c_4(x_3^3+3x_3r^2)+c_5(r^3+3rx_3^2)$, etc. In this way we can find the infinite (but countable) set of exact solutions for $M$. A more difficult step is to recognize the fact that such polynomials are only particular case of the generic solution $P=F(r+x_3)+G(r-x_3)$ were $F(.)$ and $G(.)$ are arbitrary functions of the arguments fixed in the brackets. It means that rather complicated second order equation with variable coefficients presented in (\ref{poly}) can be reduced to the D'Alambert equation if we represent the dependent variable in form (\ref{poly1}) and change the independent variables $(x_3, x_5)$ to $(x_3,\sqrt{x_5})$.

 The corresponding inverse masses and potentials  are presented in Item  7 of Table 1.

The remaining tasks, i.e.,  the constructions of exact solutions for the determining equations (\ref{Q4})-(\ref{Q8}) appear to be much more difficult problems which, however,  are solvable.

Consider  equations (\ref{Q4}). Excluding variable $N$ we obtain the following compatibility conditions for this system:
\begin{gather}\la{Q10}\begin{array}{l}-(W_\pm^2\mp\tilde r^3x_3^2)\p_{34}M+x_3W_\pm( 4x_4\p_4M-\p_{33}M)\\+(20x_4-8x_3(x_3\mp1))\p_4M+(3W_\pm+6x_3^2)\p_3M+12x_3M.\end{array}\end{gather}

This rather complicated partial differential equation of second order can be solved by the following trick. Let us choose new independent variables. To find the first of them we solve equations (\ref{Q4}) for $N=0$ and obtain the following subclass of  solutions for our problem:
\begin{gather}\la{Q11}M=
 \frac{G\left(\frac{r^2\pm1}{\tilde r}\right)}{(r^2\pm1)^2 \mp4\tilde r^2}.\end{gather}
  This solution is valid for any arbitrary function $G\left(\frac{r^2\pm1}{\tilde r}\right)$, in particular, for $G\left(\frac{r^2\pm1}{\tilde r}\right)=1$, when
  \begin{gather}\la{Q12}M=
\frac{1}{(r^2\pm1)^2 \mp4\tilde r^2}.\end{gather}
Thus it is reasonable to search for solutions with non-trivial $N$ in the form:
\begin{gather}\la{Q13}M=
 \frac{F(r,x_3)}{(r^2\pm1)^2 \mp4\tilde r^2}.\end{gather}
 Substituting this form into (\ref{Q4}) we immediately recognize that the latter equations turn to the identities provided $N=-F(r,x_3)$ and
 \begin{gather}\la{Q14} F=F\left(\frac{r^2\mp1}{x_3}\right).\end{gather}

In this way we come to the solutions presented in Item 8 of Table 1.

In complete analogy with the above one can solve equations (\ref{Q5}).
  The inverse masses and potentials generated by solutions of these equations
  are presented in Item 9 of Table 1.

  Let us consider equations (\ref{Q6}).
  Excluding $N$  we obtain the following compatibility condition for this system:
  \begin{gather}((x_4-x_3^2-1)\p_{3}\p_4+4x_3x_4\p_{4}\p_4-
  x_3\p_{3}\p_3-3\p_3+8x_3\p_4)M(x_4,x_3)=0.\la{O2}\end{gather}

  Like (\ref{poly}) equation (\ref{O2})  can be reduced to the D'Alambert equation if we chose the following  new independent and dependent variables:
   \begin{gather}\la{O3}x=r^2\pm1, \ y=x_3^2, \ \tilde M(x,y)=\sqrt{x^2\mp4y}M(\tilde r,x_3).\end{gather}
      As a result we come to the following  generic solution for (\ref{O2}):
  \begin{gather}\la{O31}M=\frac{F( x_-)+G( \tilde x_-)}{\sqrt{(r^2\pm1)^2\mp 4x_3^2}}\end{gather}
  where $x_-=\sqrt{x^2\mp 4y}+ x, \tilde x_-=\sqrt{x^2\pm 4y}- x$, $F(.)$ and $G(.)$ are arbitrary functions. The corresponding potential $V$ (\ref{Pot}) looks as:
  \begin{gather}\la{O32}V=\frac{\tilde F( x_-)+\tilde G( \tilde x_-)}{ F( x_-)+ G( \tilde x_-)}.\end{gather}

  For any fixed $F( x_-)$ and $G(\tilde x_-)$  we can solve equations (\ref{Q6}) and find functions $N=N_{F,G}$ corresponding to $M$ defined in (\ref{O31}). Unfortunately, it is seemed be impossible to represent functions $N_{F,G}$ in closed form for $F$ and $G(\tilde x_-)$ arbitrary. However, we can do it at least for some rather extended classes of $F(.)$ and $G$. In particular it is the case if $F(x_-) + G(\tilde x_-)$ is a homogeneous function of $x_3^2$ and $y$, say, for
  \begin{gather}\la{O4}F(x_-) + G(\tilde x_-)={x_-^n+(-1)^{n+1}\tilde x_-^n}=\Phi_n.\end{gather}
  Substituting (\ref{O4}) into  (\ref{O3}) and integrating the corresponding equations (\ref{me1}) we find the related  functions $M$ and $N$ in closed form for $n$ arbitrary:
  \begin{gather}\la{O5}\begin{split}&M=M_n=\frac{\Phi_n}{\sqrt{(r^2-1)^2+ 4x_3^2}},\\&N=N_n=2^nx_3^2M_{n-1}.\end{split}\end{gather}

  Thus we found the infinite (but countable) set of solutions of the determining equations (\ref{me1}) which generate integrals of motion (\ref{S013}). Let us represent explicitly some of them:
   \begin{gather}\la{O6}\begin{split}&M_1=r^2-1, \ M_{-1}=\frac1{x_3^2}\\& M_2=(r^2-1)^2+x_3^2,\\&M_3=((r^2-1)((r^2-1)^2+2x_3^2)\\&M_4=(r^2-1)^4+3(r^2-1)x_3^4+x_3^6,\\&
   M_5=(r^2-1)^5+4(r^2-1)^3x_3^2+(r^2-1)x_3^4,\\&...\\&M_n=\sum_{m\leq n}\frac{(n-m)!x^{n-2m}x_3^{2m}}{(n-2m)!2^m m!}.\end{split}\end{gather}

   Notice that linear combinations of solutions (\ref{O5}),  (\ref{O6})  also solve equations (\ref{me1}), see Item 10 of Table 1.

  The last class of symmetries which we are supposed to study is represented in equation (\ref{S015}). The corresponding determining equations (\ref{Q7}) are the most complicated, but in many aspects they are analogous to ones requested for the systems admitting integrals of motion of  type  (\ref{S013}). The related solutions are represented
 in Items  11-12 of Table 1, see preprint \cite{AG5}  for more detailed calculations. Solutions corresponding to $n^2-1<0$ which are represented in Items 13 and 14 of the same table can be obtained in analogous way.

  Thus we find all inequivalent PDM systems which, in addition to the cylindric symmetry, admit at least
  one second order scalar integral of motion and so are integrable. The inverse masses and potentials of these   systems are defined up to two arbitrary functions for pseudo scalar integrals of motion and up to four   arbitrary functions if  integrals of motion are true scalars. For some particular classes of the mentioned   arbitrary functions the number of the admitted scalar integrals of motion can be extended as it is indicated   in  Tables 3-5.

In Table 1 and the following Tables 2-5 $F(.), G(.), \tilde F(.)$ and $\tilde G(.)$ are arbitrary functions, $g(F,G)$  and  $\eta(\tilde F,\tilde G)$ are solutions of equation (\ref{Q6}), (\ref{Q7}) with given $F_1(x), F_2(\tilde x)$ and $G_1, G_2$, $c_1, c_2, ..$ are arbitrary real parameters.
  \newpage
  \begin{center}Table 1.  Inverse   masses, potentials  and scalar integrals of motion for integrable systems \end{center}
\begin{tabular}{c c c c}
\hline
\vspace{1.5mm}No&$f$&$V$&\text{Integrals of motion}\\
\hline\\

1\vspace{1.5mm}&$ \frac{\tilde r^2}{c_1 x_3 +F(\tilde r)}$&$\frac{c_2 x_3+G(\tilde r)}{c_1 x_3 +F(\tilde r)}$&$P_3L_3+\frac{c_1}2(\varphi\cdot H)-\frac{c_2}2\varphi$\\

2\vspace{1.5mm}&$\frac{\tilde r^2}{F\left(\frac{r^2-1}{\tilde r}\right)+c_1\text{arctanh}\left(\frac{r^2+1}{2x_3}\right)}$&$\frac{G\left(\frac{r^2-1}{\tilde r}\right)+c_2\text{arctanh}\left(\frac{r^2+1}{2x_3}\right)}{F\left(\frac{r^2-1}{\tilde r}\right)+c_1\text{arctanh}\left(\frac{r^2+1}{2x_3}\right)}$&$\begin{array}{c}\{L_3,(K_3+ P_3)\}\\-2c_1\left(\varphi\cdot H\right)+2c_2\varphi\end{array}$\\

3\vspace{1.5mm}&$\frac{\tilde r^2}{c_1 \arctan\left(\frac{r^2-1}{2x_3}\right)+F\left(\frac{r^2+1}{\tilde r}\right)}$&$\frac{c_2 \arctan\left(\frac{r^2-1}{2x_3}\right)+G\left(\frac{r^2+1}{\tilde r}\right)}{c_1 \arctan\left(\frac{r^2-1}{2x_3}\right)+F\left(\frac{r^2+1}{\tilde r}\right)}$&$\begin{array}{c}\{L_3,(K_3- P_3)\}
\\-2c_1\left(\varphi\cdot H\right)+2c_2\varphi
\end{array}
$\\

4\vspace{1.5mm}&$\frac{\tilde r^2}{F(\theta)-2c_1\ln(\tilde r)}$&$\frac{G(\theta)-2c_2\ln(\tilde r)}{F(\theta)-2c_1\ln(\tilde r)}$&$\begin{array}{c}DL_3+c_1(\varphi\cdot H)-c_2\varphi\end{array}$\\

5\vspace{1.5mm}&$\frac1{F(\tilde r)+G(x_3)}$&$\frac{\tilde F(\tilde r)+\tilde G(x_3)}{F(\tilde r)+G(x_3)}
$&$\begin{array}{c}P_3^2-\left(G(x_3)\cdot H\right)+\tilde G(x_3)\end{array}$\\

6\vspace{1.5mm}&$\frac{r^2}{F(\theta)+G(r)}$&$\frac{\tilde F(\theta)+\tilde G(r)}{F(\theta)+G(r)}$&$L_1^2+L_2^2-
(F(\theta)\cdot H)+\tilde F(\theta)$\\

7\vspace{1.5mm}&$\begin{array}{c}\frac{ r^2}{ F(r+x_3)+G(r-x_3)} \end{array}$&$\frac{\tilde F(r+x_3)+\tilde G(r-x_3)}{ F(r+x_3)+G(r-x_3)}$&$\begin{array}{c}\{P_3,D\}-P_nx_3P_n\\-((F(r+x_3)-G(r-x_3))\cdot H)\\+\tilde F(r+x_3)-\tilde G(r-x_3)\end{array}$\\

8\vspace{1.5mm}&$\frac{F_\pm}{F\left(\frac{r^2\pm1}{{\tilde r}}\right)-G\left(\frac{r^2\mp1}{x_3}\right)}$&$\frac{\tilde F\left(\frac{r^2\pm1}{\tilde r}\right)+\tilde G\left(\frac{r^2\mp1}{x_3}\right)}{F\left(\frac{r^2\pm1}{
{\tilde r}}\right)-G\left(\frac{r^2\mp1}{x_3}\right)}$&$\begin{array}{c}(K_3\mp P_3)^2+\tilde G\left(\frac{r^2\mp1}{x_3}\right)\\+
\left(G\left(\frac{r^2\mp1}{x_3}\right)\cdot H\right)-\hat\eta\end{array}$\\

9\vspace{1.5mm}&$\begin{array}{c}\vspace{1mm}\frac{\sqrt{F_\pm}}{F( x_\pm)+G( \tilde  x_\pm)}
\end{array}
$&$\frac{\tilde F(  x_\pm)+\tilde G( \tilde  x_\pm)}{F_1( x_\pm)+F_2( \tilde  x_\pm)}$&$\begin{array}{c}\{P_3,(K_3\pm P_3)\}+P_ax_3^2P_a\\+\left((g(F,G)\cdot H\right)-\eta((\tilde F,\tilde G)
\end{array}$\\

10\vspace{1.5mm}&$\begin{array}{c}\sum_n {c_n}F_n,\ F_n=\frac{x_\pm^n+(-1)^{n+1}\tilde x_\pm^n}{F_+}\end{array}$&$\begin{array}{c}\frac{\sum_n \tilde c_nF_n}{\sum_n c_nF_n}\end{array}$&$
\begin{array}{c}\{P_3,(K_3\pm P_3)\}+P_ax_3^2P_a\\-\left(x_3^2\sum_n  c_n 2^n F_{n-1}\cdot H\right)\\+x_3^2\sum_n \tilde c_n  2^nF_{n-1}
\end{array}$\\

11\vspace{1.5mm}&$\frac{\sqrt{z_\pm^2+2x_3^2\tilde r^2}}{F({y_\pm}
)+ G({\tilde y_\pm})}$&$\frac {\tilde F({y_\pm}
)+\tilde G({\tilde y_\pm})}{F({y_\pm}
)+G({\tilde y_\pm})}$&$\begin{array}{c}K_3^2\pm P_3^2+ 2n\{K_3,P_3\}+4P_a\tilde r^2P_a\\+\left(g(F,G)\cdot H\right)-\tilde \eta(\tilde F,\tilde G)-\hat\eta,\\ 1-n^2\leq0
\end{array}$\\

12\vspace{1.5mm}&$\begin{array}{c}\sum_m c_m\tilde F_m,\ \tilde F_m=\frac{y_\pm^m+(-1)^{m+1}\tilde y_\pm^m}{ z_\pm }\end{array}$&$\begin{array}{c}\frac{\sum_m \tilde c_m\tilde F_m}{\sum_m c_m\tilde F_m}\end{array}$&$
\begin{array}{c}K_3^2\pm P_3^2+ 2n\{K_3,P_3\}+4P_a\tilde r^2P_a\\-\left(x_3^2\sum_m  c_m 2^{m-1}\tilde F_{m-1}\cdot H\right)\\+x_3^2\sum_m \tilde c_m 2^{m-1} \tilde F_{m-1}-\hat\eta,\\ 1-n^2\leq0
\end{array}$\\

13\vspace{1.5mm}&$\frac{\sqrt{\tilde z_\pm^2\mp 2x_3^2\tilde r^2}}{F({s_\pm}
)+G({\tilde s_\pm})}$&$\frac {\tilde F({s_\pm}
)+\tilde G({\tilde s_\pm})}{F({s_\pm}
)+G({\tilde s_\pm})}$&$\begin{array}{c}K_3^2\pm P_3^2+ 2n\{K_3,P_3\}+4P_a\tilde r^2P_a\\+\left(g(F,G)\cdot H\right)-\tilde \eta(\tilde F,\tilde G)-\hat\eta, \\  n^2-1<0
\end{array}$\\

14\vspace{1.5mm}&$\begin{array}{c}\sum_m c_m\Phi_m,\ \Phi_m=\frac{s_\pm^m+(-1)^{m+1}\tilde s_\pm^m}{ z_\pm }\end{array}$&$\begin{array}{c}\frac{\sum_m \tilde c_m\Phi_m}{\sum_m c_m\Phi_m}\end{array}$&$
\begin{array}{c}K_3^2\pm P_3^2+2n\{K_3,P_3\}+4P_a\tilde r^2P_a\\-\left(x_3^2\sum_m  c_m 2^{m-1}\Phi_{m-1}\cdot H\right)\\+x_3^2\sum_m \tilde c_m 2^{m-1} \Phi_{m-1}-\hat\eta,\\ n^2-1<0
\end{array}$\\

\hline\hline
\end{tabular}
\vspace{3mm}

  In addition we use the following notations which also will be used in all tables:
\begin{gather} \la{nota}\begin{split}& F_\pm=(r^2\pm1)^2\mp4x_3^2, \ \tilde F_\pm=(r^2\pm1)^2\pm 4x_3^2,\\&\hat \eta=3(r^2- 5x_3^2), \ G_\pm=(\tilde r^2-2x_3^2) (r^2\pm1)^2+4x_3^2\tilde r^2,\\& z_\pm=
\frac{r^4\pm 1+2n(\tilde r^2-x_3^2)}{2\sqrt{2(n^2\mp1)}}, \ \ - n^2\leq 1,\\& y_\pm=\frac1{\tilde r^2}\left(\sqrt{z_\pm^2+2x_3^2\tilde r^2}+ z_\pm\right), \ \tilde y_\pm=\frac1{\tilde r^2}\left(\sqrt{z_\pm^2+2x_3^2\tilde r^2}- z_\pm\right),\\&\tilde z_\pm=\frac{r^4\pm 1+2n(\tilde r^2-x_3^2)}{2\sqrt{2(1\mp n^2)}}, \ n^2<1\\& s_{\pm}=\frac1{\tilde r^2}\left(\sqrt{\tilde z_\pm^2\mp 2x_3^2\tilde r^2}+ \tilde z_\pm\right), \ \tilde s_{\pm}=\frac1{\tilde r^2}\left(\sqrt{\tilde z^2\mp 2x_3^2\tilde r^2}- \tilde z_\pm\right),\\& x_\pm=\sqrt{(r^2\pm1)^2\mp x_3^2}+r^2\pm1, \tilde x_\pm=\sqrt{(r^2\pm1)^2\mp x_3^2}-r^2\mp1.\end{split}\end{gather}

The mysterious  term $\hat\eta$ appears as a result of the following identity:
$K_3^2=P_a\mu_8^{ab}P_b+\hat\eta$
where $\mu_9^{ab}$ is the Killing tensor (\ref{K9}) with $\lambda_9^3\neq0$ , and it is the cost paid for our desire to represent the integrals of motion via the anticommutators of operators (\ref{QQ}).

 \section{Vector integrals of motion}
The systems admitting vector integrals of motion can be classified in the way analogous to one used for the case of
scalar integrals. First, using our knowledge of the requested Killing tensors we will find the generic
form of second order integrals of motion which transform as a vectors under rotations with respect to
 the third coordinate axis. Then, using the discrete equivalence transformations we will  select the subsets of
  such  integrals with fixed parities. Then, using the continuous equivalence transformations we will specify all
  inequivalent linear combinations of the Killing tensors which should be considered. Finally we will solve
  the determining equations generated by such inequivalent Killing tensors.

  \subsection{Generic vector integrals of motion and their parity properties}

  Vector integrals of motion are generated by linear combinations of Killing tensors (\ref{K1})-(\ref{K9}) with nonzero parameters
  $\lambda^\alpha_n$ and   $\lambda^{3\alpha}_n$, where $\alpha =1, 2, n=1,...,9$. The corresponding
  integrals of motion  (\ref{QQQ}) are linear combinations of operator $Q^{(0)}$ generated by
  $\mu^{ab}_0$, i.e.,
  \begin{gather}\la{v0}Q^{(0)}=P_a g({\bf x})P_a\end{gather}
     and the following operators:
  \begin{gather}  \la{v1}  \{S_{12},S_{3\alpha}\}, \
  \{S_{12},S_{4\alpha}\}, \ \{S_{12},S_{0\alpha}\},\\  \la{v2}
  \{S_{3\alpha},S_{03}\}, \ \{S_{3\alpha},S_{43}\}, \ \{S_{3\alpha},S_{04}\}, \ \{S_{0\alpha},S_{43}\},\
   \{S_{4\alpha},S_{03}\}.
  \end{gather}

  Notice that operators (\ref{v1}) are invariant with respect to the reflections
$x_\alpha\to-x_\alpha$ while operators (\ref{v2}) change their size under this operation. Thus operators
(\ref{v1}) and (\ref{v2}) are true  and pseudo vectors respectively.

The generic vector integral of motion includes a linear combination of operators (\ref{v1}) and (\ref{v2}).
Since Hamiltonians (\ref{H1}) are invariant with respect to the space reflection, such combinations can include
either operators  (\ref{v1}) or (\ref{v2}), but never both of them. Thus it is necessary to consider the true vector and
pseudo vector integrals of motion separately, while their linear combinations can be rearrangered in accordance with the following comment.

Let us note that  pseudo vectors  (\ref{v1}) have the following property: whenever we change the sign of $x_1$, the first components of these vectors are not changed but the second components change their sign. Analogously, if we change the sign of $x_2$: the second components are kept unchanged while the first   components change their signs also. In other words, if  $V_\alpha$ is one of pseudo vectors  (\ref{v1}), then $\tilde V_\alpha=\varepsilon_{\alpha \nu}V_\nu$ (were $\varepsilon_{\alpha \nu}$ is the antisymmetric unit tensor, $\alpha$ and
$ \nu$ take the values 1 or 2) transforms as a true vector under the space reflections.

That is why  the true vectors and pseudo vectors the generic forms of the related
operator $Q^{(0)}$  (\ref{v0}) are
\begin{gather}\la{0001}Q^{(0)}_\alpha=P_ax_\alpha g({\tilde r,x_3})P_a\end{gather}
and
\begin{gather}\la{0002}\tilde Q^{(0)}_\alpha=P_a\varepsilon_{\alpha \nu}x_\nu g({\tilde r,x_3})P_a\end{gather}
correspondingly, were $\varepsilon_{\alpha \nu}$ is the antisymmetric unit tensor, $\alpha$ and
$ \nu$ take the values 1 or 2.

Like in Section 5, we can use the equivalence transformations from group SO(1,2) whose generators are $S_{03},
S_{04}$ and $S_{43}$ to simplify the integrals o motion. Let us discuss these possible simplifications.

In accordance with (\ref{K1})-(\ref{K9})  it is possible to specify five classes of vector integrals of motion, whose coefficients are zero, first, second, third or four order polynomials in $x_a$ which are fixed in the following equations (\ref{o1}), (\ref{o2}), (\ref{o3}), (\ref{o4}) and (\ref{o5}) correspondingly:
\begin{gather}\la{o1}Q^{(1)}_\alpha =P_3P_\alpha +Q^{(0)}_\alpha+x_{\alpha}\tilde\eta_1, \ \mu^{ab}=\mu^{ab}_1,\\\la{o2} \begin{split} & Q^{(2)}_\alpha=\{P_\alpha,D\}+Q^{(0)}_\alpha+x_{\alpha}\tilde\eta_2, \ \mu^{ab}=\mu^{ab}_2, \\& Q^{(3)}_\alpha=\varepsilon_{\alpha\nu} \{P_3,L_\nu\}+Q^{(0)}_\alpha+x_{\alpha}\tilde\eta_3,  \ \mu^{ab}=\mu^{ab}_2+\mu^{ab}_3, \end{split}\\ \la{o3}\begin{split}&
Q^{(4)}_\alpha=\{K_3,P_\alpha\}+ Q^{(0)}_\alpha+x_{\alpha}\tilde\eta_4,   \ \mu^{ab}=\mu^{ab}_4, \\& Q^{(5)}_\alpha=\varepsilon_{\alpha\nu} \{D, L_\nu\}+ Q^{(0)}_\alpha+x_{\alpha}\tilde\eta_5,  \ \mu^{ab}=\mu^{ab}_4+\mu^{ab}_6, \end{split}  \\ \la{o4} \begin{split}& Q^{(6)}_\alpha= \{D,K_\alpha\}+Q^{(0)}_\alpha+x_{\alpha}\tilde\eta_6,   \ \mu^{ab}=\mu^{ab}_7\\& Q^{(7)}_\alpha=\varepsilon_{\alpha\nu}\{K_3,L_\nu\} +  Q^{(0)}_\alpha+x_{\alpha}\tilde\eta_7,\ \mu^{ab}=\mu^{ab}_
7+\mu^{ab}_8,\end{split}\\\la{o5}Q^{(8)}_\alpha= \{K_3,K_\alpha\}+Q^{(0)}_\alpha+x_{\alpha}\tilde\eta_8, \ \mu^{ab}=\mu^{ab}_9\end{gather}
where $\tilde \eta_n=\tilde \eta_n(\tilde r, x_3)$ are an unknown  functions, $\mu^{ab}_1,..,\mu^{ab}_8$ are tensors (\ref{K1})-(\ref{K9}) with the following nonzero coefficients:
\begin{gather}\la{oo}\lambda^{\alpha}_n, \ \lambda^{3\alpha}_m,\ \alpha=1, 2, \ n=2, 4, 6, m=1, 3, 7, 8, 9.\end{gather}

where $\tilde \eta_n=\tilde \eta_n(\tilde r, x_3)$ are an unknown  functions, $\mu^{ab}_1,..,\mu^{ab}_8$ are tensors (\ref{K1})-(\ref{K9}) with the following nonzero coefficients:
\begin{gather}\la{oo}\lambda^{\alpha}_n, \ \lambda^{3\alpha}_m,\ \alpha=1, 2, \ n=2, 4, 6, m=1, 3, 7, 8, 9.\end{gather}

In accordance with  (\ref{o1})- (\ref{o5}), to generate vector integrals of motion we choose the special Killing tensors whose nonzero parameters are fixed in (\ref{oo}). In addition,  functions  $\eta$ and $g$ in  (\ref{Q}) and (\ref{v0}) and functions $N, K$ in (\ref{NV}) should be changed in the following manner:
\begin{gather}\la{o51}\eta\to x_\alpha\tilde \eta(\tilde r, x_3),\ N \to x_\alpha\tilde N(\tilde r,x_3), \ K\to x_\alpha\tilde K(\tilde r,x_3).\end{gather}

Let us note that operators $Q^{(8)}, Q^{(7)}$ and $Q^{(6)}$ are equivalent to $Q^{(1)}, Q^{(3)}$ and $Q^{(2)}$ respectively up to transformation (\ref{soo}) and so it is sufficient to consider only the potential integrals of motion represented in   (\ref{o1})-(\ref{o3}).
The generic vector integral of motion is a linear combination of operators (\ref{o1})-(\ref{o5}). However, such linear combination can be a priori simplified using the equivalence transformations. More exactly, we are supposed to consider the following inequivalent versions of vector integrals of motion (see \cite{AG5})
\begin{gather}\la{lc2}\begin{split}&Q=\{K_3,K_\alpha\}\pm2P_3P_\alpha+c_4\{K_3,P_\alpha\}+c_5\varepsilon_{\alpha\nu} DL_\nu+Q^{(0)}_\alpha+x_{\alpha}\tilde\eta,\\
& Q=\{(K_3\pm P_3),P_\alpha\}+c_5\varepsilon_{\alpha\nu} DL_\nu+Q^{(0)}_\alpha+x_{\alpha}\tilde\eta,\\&
Q=\{(K_3\pm P_3),P_\alpha\}+c_5\varepsilon_{\alpha\nu} DL_\nu+Q^{(0)}_\alpha+x_{\alpha}\tilde\eta,\\&
Q=P_3P_\alpha+c_5\varepsilon_{\alpha\nu} DL_\nu+Q^{(0)}_\alpha+x_{\alpha}\tilde\eta,
\\&Q=c_4\{K_3,P_a\}+c_5\varepsilon_{\alpha\nu} DL_\nu+Q^{(0)}_\alpha+x_{\alpha}\tilde\eta\end{split}\\
\la{lc6}\begin{split}&Q=c_1\{D,(K_\alpha\pm P_\alpha)\}+c_2\varepsilon_{\alpha\nu}\{L_\nu,(K_3\pm P_3)\}+Q^{(0)}_\alpha+x_{\alpha}\tilde\eta,\\
&Q=c_1\{D,P_\alpha\}+c_2\varepsilon_{\alpha\nu}\{L_\nu,(K_3\pm P_3)\}+c_3\varepsilon_{\alpha\nu}\{L_\nu,P_3\}+Q^{(0)}_\alpha+x_{\alpha}\tilde\eta,\\
&Q=c_1\{D,(K_\alpha\pm P_\alpha)\}+c_2\varepsilon_{\alpha\nu}\{L_\nu, P_\alpha\}+Q^{(0)}_\alpha+x_{\alpha}\tilde\eta,\\&
Q=c_1\{D, P_\alpha\}+c_2\varepsilon_{\alpha\nu}\{L_\nu, P_\alpha\}+Q^{(0)}_\alpha+x_{\alpha}\tilde\eta.\end{split}
\end{gather}

Thus to classify the PDM systems with cylindric symmetry which admit vector integrals of motion we are supposed to solve the determining equations (\ref{me1}), (\ref{me2}) which correspond to the symmetries represented in equations (\ref{o1}) - (\ref{o3}) and (\ref{lc2}), (\ref{lc6}).

\subsection{Selected calculations}

The next step is to solve the the determining equations indicated in the above.

For the symmetry specified in (\ref{o1}) the determining equations (\ref{me1}) have especially simple form. Namely, substituting the Killing tensor $\mu^{3\alpha}_1$ for $\alpha=1$ into (\ref{me1}) we come to the following system:
\begin{gather}\la{a1}\begin{split}& \p_3M(\tilde r,x_3)=\p_1(x_1\tilde N(\tilde r,x_3)),\\&
\p_2\tilde N(\tilde r,x_3)=0, \ \p_1M(\tilde r,x3)=\p_3(x_1\tilde N(\tilde r,x_3))\end{split}\end{gather}
which is easy solvable. Its generic solution is:
\begin{gather}\la{a2}M=c_1r^2+c_2x_3+c_3, \ \tilde N=-c_1x_3-c_2.\end{gather}

In contrast with the case of scalar integrals of motion (refer to Table 1) the obtained solution do not include arbitrary functions but only arbitrary parameters $c_1, c_2$ and $c_3$.

 The reason of the reduced freedom in the arbitrary elements $M$ lies in the fact that the considered systems by definition admit as minimum two integrals of motion which are the components of the bivector (\ref{o1}). We have found the first component, i.e, choose $\alpha =1$ in (\ref{o1}) and (\ref{o51}) but the second one can be obtained simple by the changes $P_1\to P_2, x_1\to x_2$.

 Moreover, it happens that our system admits two more integrals of motion and is maximally superintegrable. Its integrals of motion together with the admissible potential are represented in Item 4 of Table 5.

Consider now  integral of motion $Q^{(2)}_a$ presented in (\ref{o2}) where $Q^{(0)}_\alpha =-P_ax_\alpha P_a $ and $\alpha=1$. The related determining equations (\ref{me1}) have the following form:
\begin{gather}\la{a4}\begin{split}&2\tilde r\p_4M(\tilde r^2,x_3)+x_3\p_3M(\tilde r^2,x_3)+M(\tilde r^2,x_3) +N(\tilde r^2,x_3)=0,\\&2x_3\p_4M(\tilde r^2,x_3)-\p_3M(\tilde r^2,x_3)+\p_3N(\tilde r^2,x_3)=0, \ \p_4N(\tilde r^2,x_3)=0\end{split}\end{gather}
and are easy solvable also. Their generic solution is:
\begin{gather}\la{ a5}M=c_1+\frac{c_2}{x_3^2}, +\frac{c_3}{\tilde r}, \  N=-c_1+\frac{c_2}{x_3^2}.\end{gather}
The corresponding inverse mass and potential are presented in Item 6 of Table 4 where one more integral of motion is indicated also. Thus the considered system is superintegrable.

The next (and the last) example we consider is the system admitting the last symmetry presented in (\ref{lc2}), i.e., $ c_1Q^{(4)}_\alpha +c_2Q^{(5)}_\alpha. $ Setting  $\alpha=1$ we come to the following determining equations (\ref{me1}):
\begin{gather*}(c_2-c_1)(x_4-x_3^2)\p_3M+(2c_1-c_2)\p_4M+2(2c_1-c_2)x_3M-2x_4\p_4N-N=0,\\
2(c_2-2c_1)x_3x_4\p_4M+((c2-c1)x_3^2+c_1x_4)\p_3M+2(c_2-2c_1))x_3\p_3M+N=0,\\
(c_1-c_2)(x_3^2-x_4)-\frac12\p_3Nc_2x_3\p_3M+c_2M=0.
\end{gather*}
Excluding $N$ we come to the following compatibility conditions of the system presented above:
\begin{gather*}(c_2-2c_1)x_3x_4\p_{34}M+\frac12(c_2-c_1)(x_3^2+c_1x_4)\p_{34}M+
\frac32(c_2-2c_1)\p_4M+\frac14(c_2+2c_1)\p_3M=0,\\
(c_1-c_2)(x_4-x_3^2)\p_{44}M+\frac32\left(c_2-\frac23\right)x_3\p_{34}M+\left(\frac52c_2-2c_1\right)\p_4M-
\frac14\p_{33}M=0.
\end{gather*}

This system has two special and one regular solutions. Namely, for $c_1=2c_2$ and $c_1=c_2$ we obtain:
\begin{gather}\la{ss1}M=c_3r\end{gather}
and
\begin{gather}\la{ss2}M=\frac{c_3x_3}{r^2}+F(r)\end{gather}
respectively, where $F(r)$ is an arbitrary function. If parameters $c_1$ and $c_2$ do not satisfy the conditions presented above, the solution is
\begin{gather}\la{ss3}M=\frac{c_3}{r}.\end{gather}

The system with whose mass is given in (\ref{ss2}) is maximally superintegrable, see Item 1 of Table 5. The solutions (\ref{ss1}) and (\ref{ss3}) correspond to the rotationally invariant systems which admit as minimum the three parametric Lie group. Such systems are completely classified in paper \cite{N1}  and we will not discuss them here.

Solving step by step the determining equations corresponding to the remaining symmetries (\ref{o3}), (\ref{o4}), (\ref{o5}),  (\ref{lc2}), (\ref{lc6}) we find all inequivalent PDM systems with cylindric symmetry, which admit vector integrals of motion. The obtained results are presented in Tables 2-5.

\newpage

\begin{center}Table 2.  Inverse   masses, potentials, vector and tensor  integrals of motion for integrable PDM systems \end{center}

\begin{tabular}{c c c c}
\hline
\vspace{1.5mm}No&$f$&$V$&\text{Integrals of motion}\\
\hline\\

1\vspace{1 mm}&$\frac{x_3^2}{c_1
+c_2(r^2\pm1)F_\pm^\frac12+c_3x_3^2 (r^2\mp1)^{-2}}$&$\frac{c_4
+c_5(r^2\pm1)F_\pm^\frac12+c_6x_3^2 (r^2\mp1)^{-2}}{c_1
+c_2(r^2\pm1)F_\pm^\frac12+c_3x_3^2 (r^2\mp1)^{-2}}$&$\begin{array}{c}\{(K_3\pm P_3),L_\alpha\}+3\varepsilon_{\alpha\nu}x_\nu\\

+\frac{2\varepsilon_{\alpha\nu}x_{\nu}(c_4(r^2\pm1)+
c_5(r^2\mp1)^2)}{x_3^2}\\-\left(\frac{2\varepsilon_{\alpha\nu}x_{\nu}(c_1(r^2\pm1)+
c_2(r^2\mp1)^2)}{x_3^2}\cdot H\right)
\end{array}$\\

\hline

2\vspace{1 mm}&$\frac{x_3^2\tilde F_\pm }{
c_2(r^2\mp1)+\tilde F_\pm(c_3x_3^2 (r^2\pm1)^{-2}+c_1)}$&$\frac{c_5(r^2\mp1)+\tilde F_\pm(c_6x_3^2 (r^2\pm1)^{-2}+c_4)}{c_2(r^2\mp1)+\tilde F_\pm(c_3x_3^2 (r^2\pm1)^{-2}+c_1)}$&$\begin{array}{c}\{(K_\alpha\pm P_\alpha),D\}-15x_\alpha\\\mp\varepsilon_{\alpha\nu}
\{L_3,(K_\nu\mp P_\nu)\}\pm3x_\alpha\\
-2 x_\alpha\left(c_5\frac{r^2\mp1}{(r^2\pm1)^2}\mp\frac{2c_6}{\tilde F_\pm}\right)\\
+2\left(x_\alpha\left(c_2\frac{r^2\mp1}{(r^2\pm1)^2}\mp\frac{2c_3}{\tilde F_\pm}\right)\cdot H\right)\\
\end{array}$\\

\hline

3\vspace{1 mm}&$\frac{x_3^6}{c_1(x_3^2 +4\tilde r^2)+c_2x_3^4+c_3x_3^6}$&$\frac{(c_4x_3^2 +4\tilde r^2)+c_5x_3^4+c_6x_3^6}{c_1(x_3^2 +4\tilde r^2)+c_2x_3^4+c_3x_3^6}$&$\begin{array}{c}\{P_\alpha,D\}+\varepsilon_{\alpha\nu} \{L_3,P_\nu\}\\
+2\left(\frac{x_\alpha(c_1-c_3x_3^4)}{x_3^4}\cdot H\right)\\-2\frac{x_\alpha(c_4-c_6x_3^4)}{x_3^4}\end{array}$\\

\hline

4\vspace{1 mm}&$\frac{r(r^2\pm1)^2x_3^2}{c_1r(r^2\pm1)^2+c_2rx_3^2+c_3(r^2\mp1)x_3^2}$&$
\frac{c_4r(r^2\pm1)^2+c_5rx_3^2+c_6(r^2\mp1)x_3^2}{c_1r(r^2\pm1)^2+c_2rx_3^2+c_3(r^2\mp1)x_3^2}
$&$\begin{array}{c}\{D,(K_\alpha\pm P_\alpha)\}-\frac{2x_\alpha c_5(r^2\mp1)}{(r^2\pm1)^2}\\-
\frac{x_\alpha c_6((r^2\mp1)\mp4r^2)}{r(r^2\pm1)^2}-15x_\alpha\\+\left(\frac{2x_\alpha c_2((r^2\mp1)}{(r^2\pm1)^2}\cdot H\right)\\
+\left(\frac{x_\alpha c_3((r^2\mp1)^2+4r^2)}{r(r^2\pm1)^2}\cdot H\right)\end{array}$\\

\hline

5\vspace{1.5mm}&$\frac{ x_3^2}{x_3^2F(r)+c_1}$&$\frac{x_3^2G(r)+c_2}{x_3^2F(r)+c_1}$&$
\begin{array}{c}\{L_1,L_2\}+\left(\frac{c_1x_1x_2}{x_3^2}\cdot H\right)-\frac{c_2x_1x_2}{x_3^2},\\L_1^2-L_2^2+\left(\frac{c_1(x_1^2-x_2^2)}{2x_3^2}\cdot h\right)-\frac{c_2(x_1^2-x_2^2)}{2x_3^2}\end{array}$\\

\hline

6\vspace{1.5mm}&$\frac{ (r^4-1)^2}{(r^2\pm1)^2F\left(\frac{r^2\mp
1}{x_3}\right)-c_1 r^2}$&$\frac{ (r^2\pm1)^2G\left(\frac{r^2\mp1}{x_3}\right)+ {c_2} r^2}{(r^2\pm1)^2F\left(\frac{r^2\mp
1}{x_3}\right)-c_1 r^2}$&$\begin{array}{c}(K_1\pm P_1)^2-(K_2\pm P_2)^2\\+15(x_1^2-x_2^2)\\+c_1\left(\frac{x_1^2-x_2^2}{(r^2\pm1)^2}\cdot H\right)+ {c_2}\frac{x_1^2-x_2^2}{(r^2\pm1)^2},\\\{(K_1\pm P_1),(K_2\pm P_2)\}\\+{c_2}\frac{x_1x_2}{(r^2\pm1)^2}+15x_1x_2
\end{array}$\\

\hline

7\vspace{1.5mm}&
$\frac{(r^2\pm1)^2F_{\pm}^2}{c_1G^{\pm}+2c_2x_3^2(r^2\pm 1)+c_3F_{\pm}^2}$&$\frac{c_4G^{\pm}+2c_5x_3^2(r^2\pm 1)+2c_6F_{\pm}^2}{c_1G^{\pm}+2c_2x_3^2(r^2\pm 1)+c_3F_{\pm}^2}$&$\begin{array}{c}\{(K_1\pm P_1),L_1\}\\-\{(K_2\pm P_2),L_2\}\\
+2\left(x_1x_2(\frac{c_1x_3(r^2\pm1)}{F_\pm^2}+c_2)\cdot H\right)\\-2x_1x_2(\frac{c_4x_3(r^2\pm1)}{F_\pm^2}+c_5),\\\{(K_1\pm P_1),L_2\}\\+\{(K_2\pm P_2),L_1\}\\+\left((x_1^2-x_2^2)(\frac{c_1x_3(r^2\pm1)}{F_\pm^2}+c_2)\cdot H\right)\\-(x_1^2-x_2^2)(\frac{c_4x_3(r^2\pm1)}{F_\pm^2}+c_5)
\end{array}$\\

\hline\hline

\end{tabular}

\vspace{3mm}

\section{Tensor integrals of motion}

The last class of symmetries which is supposed to be considered are the tensor integrals of motion. The corresponding Killing tensors are given by equations (\ref{K1}), (\ref{K3}), (\ref{K6}), (\ref{K8}) and (\ref{K9}) where where the  indices $a, b$ of the  nonzero parameters $\lambda_k^{ab}$ independently take the values $1$ and $2$. In other words, we have to  consider exactly five linearly independent tensors and their linear combinations. The corresponding integrals of motion include operators (\ref{v0}) and the following bilinear combinations:
\begin{gather}\la{T1}P_\alpha P_\nu, \ \{K_\alpha K_\nu\}, \ \{P_\alpha K_\nu\}\end{gather}
and
\begin{gather}\la{T2}\{P_\alpha,L_\nu\}, \ \{K_\alpha L_\nu\}\end{gather}
were $\alpha$ and $\nu$ take the values 1 and 2.

Notice that operators (\ref{T1}) are true tensors while operators (\ref{T2}) change their signs under the reflection and so are pseudo tensors. It means that we have to consider linear combinations of
symmetries (\ref{T1}) and  (\ref{T2}) separately.

One more note is that the traceless symmetric  tensor $Q^{\alpha\nu}$ in two dimensions has exactly two linearly independent components, i.e., $Q^{12}$ and  $Q^{11}-Q^{22}$. We will work with the first components, i.e., $Q^{12}$ and will omit the top index $12$.

 \subsection{True tensor integrals of motion}

Up to equivalence transformation (\ref{IT}) we can specify the following inequivalent linear combinations of true scalars (\ref{ss1}) and the universal block (\ref{v0}):
\begin{gather}\la{ss4}Q_1=Q^{(0)}_1+P_{1}P_2+\eta_1,\\
\la{ss5} Q_2=Q^{(0)}_2\{K_1,K_2\}\pm 2P_1P_2+\eta_2,\\
\la{ss6}Q_3=Q^{(0)}+\{P_{1},P_2\pm K_2\}+\eta_3,\\\la{ss7}Q_4=Q^{(0)}+\{P_{1}\pm K_1,P_2\pm K_2\}+\eta_4.
\end{gather}

The determining equations (\ref{me1}) for symmetry (\ref{ss4}) are simple and have the following form:
\begin{gather}\la{ss8}\p_2 M+\p_1 N=0, \ \p_1 M+\p_2 N=0, \ \p_3 N=0.\end{gather}
Taking into account that the generic form of functions $M$ and $N$ is given by the following equation
\begin{gather*}M=M(\tilde r, x_3), \ N=x_1x_2\tilde N(\tilde r,x_3)\end{gather*}
the system (\ref{ss8}) is easy integrated and solved by the following functions:
\begin{gather*}M=c_1r^2+F(x_3),  \ N=c_1x_1x_2.\end{gather*}
The corresponding PDM system appears to be superintegrable, see Item 3 of Table 4.

Let us consider one more integral of motion which is specified  in (\ref{ss7}). Denoting $N=x_1x_2\tilde  N(\tilde r^2,x_3)$ we come to the following corresponding equations (\ref{me1}):
\begin{gather*}\p_4((x_4-x_3^2\pm 1)^2M)+x_3(x_4-x_3^2\pm 1)\p_3M+\frac12\p_4(x_4\tilde N)=0,\\x_3(x_4-x_3^2-1)\p_4M+\p_3(x_3^2M)+\frac18\p_3\tilde N\end{gather*}
which are perfectly solved by the following functions:
\begin{gather*}M=\frac{(r^2\pm1)F\left(\frac{r^2\mp1}{x_3}\right)-c_1r^2}{(r^4-1)^2}, \ \tilde N=\frac{c_1x_1x_2}{(r^2\pm1)^2}.\end{gather*}

The obtained results are represented in Item 6 of Table 2.

We will not discuss the determining equations for symmetries (\ref{ss5}) and (\ref{ss6}) but mention that both of them can be solved exactly also. However, for symmetry (\ref{ss5}) the related solutions are functions of $x_3$ only, but for symmetry (\ref{ss6}) we obtain the mass function which depends only on $r$. Such systems are out of the scop of the present paper since they were discussed in papers \cite{AG} and \cite{AG1}.

\subsection{Pseudo tensor integrals of motion}

Such integrals should include linear combinations of the terms presented in (\ref{T2}). Up to the equivalence we can restrict ourselves to two types of such linear combinations:
\begin{gather}\la{ss10}Q_6=Q^{(0)}+\{P_1,L_2\}+\{P_2,L_1\}+\eta_6\end{gather}
and
\begin{gather}\la{ss11}Q_7=Q^{(0)}+\{(P_1\pm K_1),L_2\}+\{(P_2\pm K_2),L_1\}+\eta_7.\end{gather}

Let us denote $N=(x_1^2-x_2^2)\tilde N(r,x_3)$ then symmetry (\ref{ss10}) generates the following version of the determining equations (\ref{me1}):
\begin{gather*} 2(x_1^2-x_2^2)\p_4\tilde N+4x_3\p_4 M-\p_3M+2\tilde N=0,\\2\p_4M-\p_3 N=0.\end{gather*}
which are solved by the following functions:
\begin{gather*}M=c_1+c_2(\tilde r^2+4x_3^2+2c_3x_3), \tilde N=-2c_2x_3-c_3.\end{gather*}
The corresponding PDM system appears to be maximally superintegrable, see Item 2 of Table
5.

The next (and the last in this section) symmetry which we consider is specified in (\ref{ss11}). The corresponding determining equations (\ref{me1}) are reduced to the following system:
\begin{gather*}x_3(x_4-x_3^2-1)\p_4M+\frac14(1+3x_3^2-x_4)\p_3M-\frac12(x_4\tilde N)+x_3M=0,\\(x_4+3x_3^2+1)\p_3M-4x_3(x_3^2+1)\p_4M+4x_3M-2\tilde N,\\(x_4-3x_3^2-1)\p_4M+2\p_3(x_3M)+\frac12\tilde N=0\end{gather*}

which is solved by the following functions:
\begin{gather*}M=c_1\frac{(\tilde r^2-2x_3^2) (r^2\pm1)^2+4x_3^2\tilde r^2}{(r^2\pm1)^2((r^2\pm1)^2\pm4x_3^2)}+2c_2\frac{x_3^2}{(r^2\pm1)((r^2\pm1)^2\pm4x_3^2)}
+\frac{c_3}{(r^2\pm1)^2},\\\tilde N=c_1\frac{x_3}{((r^2\pm1)^2\pm4x_3^2)}+c_2.
\end{gather*}

The related PDM system is integrable, see Item 7 of Table 2.

\section{Superintegrable systems}

Thus we have specified all nonequivalent cylindrically invariant PDM systems which admit a second order integral of motion. By definition such systems admit two integrals of motion one of which is the generator of rotations  around the third coordinate axis, and so they are integrable. All the systems admitting two (but no more) integrals of motion are presented in Tables 1 and 2.

However, some of the found systems automatically admit more than two integrals of motion. In addition, for some particular versions of the arbitrary elements, i.e., arbitrary functions and integration constants we also can find additional integrals of motion. The related systems are superintegrable or even maximally superintegrable.

The next (and final) step of our analysis is the specifications of just these systems. To do it wee supposed to solve the extended systems of the determining equations (\ref{me1}) corresponding to the inequivalent pairs of integrals of motion.

The same speculations with the inequivalent triplets of integrals of motion and the related mass functions make it possible to specify the inequivalent maximally superintegrable systems.

To specify the PDM systems  as minimum two second order scalar integrals of motion we have to search for such cases when arbitrary elements $M$ and $N $ satisfy more than one system of the determining equations specified in (\ref{s001}),  (\ref{002}), (\ref{003}), (\ref{006}) and (\ref{Q1})-(\ref{Q8}).

Solving step by step all pairs of the mentioned equations we find the superintegrable systems admitting scalar integrals of motion. The obtained results are presented in Table 3
and the following equations (\ref{last4})-(\ref{last6}).

Summarizing, we find all inequivalent  superintegrable PDM systems admitting scalar integrals of motion. The number of such systems  appears to be rather extended, see Table 3 and the following formulae (\ref{last4})-(\ref{last5}). The latter formulae collect such expressions  which are too cumbersome to be placed in a table.

The next class of superintegrable PDM systems which we consider are those ones which admit both the scalar and vector or tensor integrals of motion. To specify such systems we have to verify the consistence of the systems of the determining equations corresponding to the combinations of symmetries.  In this way we obtain the results presented in Table 4.

The final step is to classify the maximally superintegrable system. It can be done in the same way as in the superintegrable case. The classification results are presented in Table 5 which includes two parts.

At this point the classification of cylindrically invariant PDM system admitting second order integrals of motion has been completed.
\newpage

\begin{center}Table 3.  Inverse   masses, potentials  and  scalar integrals of motion for superintegrable systems \end{center}
\begin{tabular}{c c c c}
\hline

\vspace{1.5mm}No&$f$&$V$&\text{Integrals of motion}\\
\hline\\

1\vspace{1.5mm}&$\frac{\tilde r^2}{\tilde r^2(c_1+2c_2x_3+c_3(\tilde r^2+4x_3^2))+c_4 }$&$\frac{\tilde r^2(c_5+2c_6x_3+c_7(\tilde r^2+4x_3^2))+c_8 }{\tilde r^2(c_1+2c_2x_3+c_3(\tilde r^2+4x_3^2))+c_4 }$&$\begin{array}{c}P_3^2+2c_6x_3+4c_7x_3^2+c_4\frac{x_3}{\tilde r^2}\\-\left((2c_2x_3+4c_3x_3^2)\cdot H\right),\\\{P_3,D\}-P_nx_3P_n\\+\left((-c_1x_3-c_2(2x_3^2+\tilde r^2)\right.\\\left.-c_3x_3(4x_3^2+3\tilde r^2))\cdot H\right)\\+c_5x_3+c_6(2x_3^2+\tilde r^2)\\-c_8\frac{x_3}{\tilde r^2}+c_7x_3(4x_3^2+3\tilde r^2)\end{array}$\\

\hline

2\vspace{1.5mm}&$\frac{r\tilde r^2}{c_1r\tilde r^2+c_2 r+c_3 x _3+c_7\tilde r^2}$&$\frac{c_4r\tilde r^2+c_5 r+c_6 x _3+c_8\tilde r^2}{c_1r\tilde r^2+c_2 r+c_3 x _3+c_7\tilde r^2}$&$\begin{array}{c}\{P_3,D\}-P_nx_3P_n\\+c_2\left(\frac{c_2x_3+c_3r}{\tilde r^2}\cdot H\right)\\-\frac{c_5x_3+c_6r}{\tilde r^2},\\L_1^2+L_2^2\\-
(\frac{r^2}{\tilde r^2}(c_2+c_3\frac{x_3}r)\cdot H)\\+(c_5+c_6\frac{x_3}r)\frac{r^2}{\tilde r^2}\end{array}$\\
\hline

3\vspace{1.5mm}&$\frac{\tilde r^2x_3^2}{c_1 x_3^2+ {c_2}\tilde r^2+{c_3} x_3^2\tilde r^2+{c_4} r^2\tilde r^2x_3^2}$&$\frac{{c_5} x_3^2+{c_6}\tilde r^2+{c_7} x_3^2\tilde r^2+{c_8} r^2\tilde r^2x_3^2}{c_1 x_3^2+ {c_2}\tilde r^2+{c_3} x_3^2\tilde r^2+{c_4} r^2\tilde r^2x_3^2}$&$\begin{array}{c}\{P_3,K_3\}+\frac{2{c_6}\tilde r^2}{x_3^2}-2\omega x_3^2\\+2\left(({c_3} x_3^2- {c_2}\frac{\tilde r^2}{x_3^2})\cdot H\right),\\P_3^2-\left((\frac {c_2}{x_3^2}+{c_4} x_3^2)\cdot H\right)\\+\frac{c_6}{x_3^2}+{c_8} x_3^2\end{array}$\\

\hline

4\vspace{1.5mm}&$\frac{\tilde r^2 x_3^2}{c_1 \tilde r^2+x_3^2(c_3 -2 {c_2} \ln(\tilde r))}$&$\frac{ x_3^2(2c_5\ln(\tilde r)+c_6)+c_4 \tilde r^2}{c_1 \tilde r^2+x_3^2(c_3 -2 {c_2} \ln(\tilde r))}$&$\begin{array}{c}P_3^2-\left((\frac {c_1}{x_3^2}\cdot H\right)+\frac{c_4}{x_3^2},\\DL_3+ {c_2}(\varphi\cdot H)-c_5\varphi\end{array}$\\

\hline

5\vspace{1.5mm}&$\frac{\tilde r^2x_3^2\sqrt{F_\pm}}{c_1x_3^2\sqrt{F_\pm}+c_2r^2\sqrt{F_\pm}+c_3\tilde r^2(r^2\pm1)}$&$
\frac{c_4x_3^2\sqrt{F_\pm}+c_5r^2\sqrt{F_\pm}+c_6\tilde r^2(r^2\pm1)}{c_1x_3^2\sqrt{F_\pm}+c_2r^2\sqrt{F_\pm}+c_3\tilde r^2(r^2\pm1)}$&$\begin{array}{l}(K_3\pm P_3)^2+\frac{F_\pm (c_5+c_6(r^2\pm1))}{x_3^2}\\-
\left(\frac{F_\pm (c_2+c_3(r^2\pm1))}{x_3^2}\cdot H\right)-\hat\eta,\\\{P_3,(K_3\pm P_3)\}+c_5\frac{(\tilde r^2\pm1)}{x_3^2}\\+c_6\frac{x_3^2(\tilde r^2\mp1)+(\tilde r^2\pm1)^2}{x_3^2\sqrt{F_\pm}}-\left(c_2\frac{(\tilde r^2\pm1)}{x_3^2}\cdot H\right)\\'
-\left(c_3\frac{x_3^2(\tilde r^2\mp1)+(\tilde r^2\pm1)^2}{x_3^2\sqrt{F_\pm}}\cdot H\right)\end{array}$\\

\hline

6\vspace{1.5mm}&$\frac{\tilde r^2x_3^2\sqrt{F_\pm}}{c_1x_3^2\sqrt{F_\pm}+c_3r^2\sqrt{F_\pm}+c_3(r^2\pm1)\tilde r^2}$&$\frac{c_4x_3^2\sqrt{F_\pm}+c_5r^2\sqrt{F_\pm}+c_6(r^2\pm1)\tilde r^2}{c_1x_3^2\sqrt{F_\pm}+c_2r^2\sqrt{F_\pm}+c_3(r^2\pm1)\tilde r^2}$&$\begin{array}{c}K_3^2-P_3^2-\hat\eta
+\frac{c_5(r^4-1)}{x_3^2}\\+\frac{c_6(r^2\mp1)(r^4\pm1+2\tilde r^2)}{x_3^2\sqrt{F_\pm}}\\-\left(\frac{c_2(r^4-1)}{x_3^2}\cdot H\right)\\-\left(\frac{c_3(r^2\mp1)(r^4\pm1+2\tilde r^2)}{x_3^2\sqrt{F_\pm}}\cdot H\right),\\(K_3\pm P_3)^2-\hat\eta+\frac{c_5(r^2\pm1)}{x_3^2}\\+\frac{c_6(r^2\pm1)^2\sqrt{F_\pm}}{x_3^2}
-\left(\frac{c_2(r^2\pm1)}{x_3^2}\cdot H\right)\\-\left(\frac{c_3(r^2\pm1)^2\sqrt{F_\pm}}{x_3^2}\cdot H\right)\end{array}$
\\

\hline\hline
\end{tabular}

\begin{gather}
\la{last4}\begin{split}
&f=\frac{\tilde F_\pm^2(r^2\mp1)^2\tilde r^2}{c_1(r^2\mp1)^2\tilde F_\pm^2+c_2x_3\tilde r^2(r^4 \pm1)(r^2\mp1)+c_3\Phi_\pm+c_4\Lambda_\pm},\\&V=\frac{c_5(r^2\mp1)\tilde F_\pm^2+c_6x_3\tilde r^2(r^4 \pm1)(r^2\mp1)+c_7\Phi_\pm+c_8\Lambda_\pm}{c_1(r^2\mp1)\tilde F_\pm^2+c_2x_3\tilde r^2(r^4 \pm1)(r^2\mp1)+c_3\Phi_\pm+c_4\Lambda_\pm},\\&\Phi_\pm=(r^2\pm1)^2((r^2\mp1)^2\pm r)+4x_3(4r^2-\tilde r^2),\\&\Lambda_\pm=(\tilde r^2+4x_3^2)(r^2\pm1)^2-4x_3^2\tilde r^2,\\&(K_3\pm P_3)^2-\hat\eta-\frac1{\tilde F_\pm^2}(c_6x_3(r^2\pm1)F_-+4(c_8-3c_7)x_3^2(r^2\pm1)^2)
\\&+\left(\frac1{\tilde F_\pm^2}(c_2x_3(r^2\pm1)F_\pm+4(c_4-3c_3)x_3^2(r^2\pm1)^2
\cdot H\right),\\&\{D,(K_3\mp P_3)\}-15x_3-\frac{c_6F_\pm}{8\tilde F_\pm^2}+\left(-\frac{c_2F_\pm}{8\tilde F_\pm^2}\cdot H\right)\\&-\frac{c_7x_3(r^2\mp1)^4+
(c_8-2c_7x_3)((r^2\pm1)^2+2\tilde r^2)(2r^2-\tilde r^2)(r^2\pm1)}{\tilde F_\pm^2(r^2\mp1)^2}\\&
+\left(\frac{c_3x_3(r^2\mp1)^4+
(c_4-2c_3x_3)((r^2\pm1)^2+2\tilde r^2)(2r^2-\tilde r^2)(r^2\pm1)}{\tilde F_\pm^2(r^2\mp1)^2}\cdot H\right);
\end{split}\\
\la{last6}\begin{split}
&f=\frac{\tilde F_\pm^2(r^2\mp1)^2\tilde r^2}{c_1(r^2\mp1)^2\tilde F_\pm^2+c_2x_3\tilde r^2(r^4 \pm1)(r^2\mp1)+c_3\Phi_\pm+c_4\Lambda_\pm},\\&V=\frac{c_5(r^2\mp1)\tilde F_\pm^2+c_6x_3\tilde r^2(r^4 \pm1)(r^2\mp1)+c_7\Phi_\pm+c_8\Lambda_\pm}{c_1(r^2\mp1)\tilde F_\pm^2+c_2x_3\tilde r^2(r^4 \pm1)(r^2\mp1)+c_3\Phi_\pm+c_4\Lambda_\pm},\\&\Phi_\pm=(r^2\pm1)^2((r^2\mp1)^2\pm r)+4x_3(4r^2-\tilde r^2),\\&\Lambda_\pm=(\tilde r^2+4x_3^2)(r^2\pm1)^2-4x_3^2\tilde r^2,\\&Q_1=(K_3\pm P_3)^2-\hat\eta-\frac1{\tilde F_\pm^2}(c_6x_3(r^2\pm1)F_-+4(c_8-3c_7)x_3^2(r^2\pm1)^2)
\\&+\left(\frac1{\tilde F_\pm^2}(c_2x_3(r^2\pm1)F_\pm+4(c_4-3c_3)x_3^2(r^2\pm1)^2
\cdot H\right),\\&Q_2=\{D,(K_3\mp P_3)\}-15x_3-\frac{c_6F_\pm}{8\tilde F_\pm^2}+\left(-\frac{c_2F_\pm}{8\tilde F_\pm^2}\cdot H\right)\\&-\frac{c_7x_3(r^2\mp1)^4+
(c_8-2c_7x_3)((r^2\pm1)^2+2\tilde r^2)(2r^2-\tilde r^2)(r^2\pm1)}{\tilde F_\pm^2(r^2\mp1)^2}\\&
+\left(\frac{c_3x_3(r^2\mp1)^4+
(c_4-2c_3x_3)((r^2\pm1)^2+2\tilde r^2)(2r^2-\tilde r^2)(r^2\pm1)}{\tilde F_\pm^2(r^2\mp1)^2}\cdot H\right).
\end{split}\end{gather}

\begin{gather}
\la{last1}\begin{split}&f=\frac{(r^4\pm1)^2x_3^2\tilde r^2}{x_3^2(c_1(r^4\mp1)^2\tilde r^2+c_3\tilde r^2(r^4\pm1)+c_4r^2\tilde r^2)+c_2(r^4\pm 1\mp 2r^2\tilde r^2)\tilde r^2},\\&V=\frac{x_3^2(c_5(r^4\mp 1)^2\tilde r^2+c_7\tilde r^2(r^4\pm 1)+c_8r^2\tilde r^2)+c_6(r^4\pm1\mp2r^2\tilde r^2)\tilde r^2}{x_3^2(c_1(r^4\mp1)^2\tilde r^2+c_3\tilde r^2(r^4\pm1)+c_4r^2\tilde r^2)+c_2(r^4\pm1\mp2r^2\tilde r^2)\tilde r^2},\\&Q_1:=K_3^2\pm P_3^2-\hat\eta-\frac{(1\pm r^4)(c_8x_3^4+c_6((r^4\mp1)^2\pm4x_3^4)+2c_7r^2x_3^4)}{x_3^2(r^4\mp1)^2}\\
&+\left(\frac{(1\pm r^4)(c_4x_3^4+c_2((r^4\mp1)^2\pm 4x_3^4)+2c_3r^2x_3^4)}{x_3^2(r^4\mp1)^2}\cdot H\right),\\
&Q_2=\{K_3,P_3\}+\frac{c_6(\tilde r^2(r^4\mp1)^2\pm2x_3^4r^2)-c_7(r^4\pm1)x_3^4-2c_8r^2x_3^4}{2x_3^2(r^4\mp1)^2}\\
&-\left(\frac{c_2(\tilde r^2(r^4\mp1)^2-2x_3^4r^2)-c_3(r^4\pm1)x_3^4-2c_4r^2x_3^4}{2x_3^2(r^4\mp1)^2}\cdot H\right);\end{split}\end{gather}\begin{gather}
\la{last2}\begin{split}&f=\frac{\tilde r^2(r^2\pm1)^2\sqrt{F_\pm}}{\sqrt{F_\pm}(c_1(r^2\pm1)^2+c_2(r^2\pm1)^2-2\tilde r^2))+c_3(r^4-1)(r^2\pm1)+c_4x_3\tilde r^2},\\
&V=\frac{c_3\tilde F_\pm^2+c_4x_3\tilde r^2(r^2\mp1)}{c_1\tilde F_\pm^2+c_2x_3\tilde r^2(r^2\mp1)},\\
&Q_1=(K_3\pm P_3)^2-\hat\eta+\frac{8c_6x_3^2+c_8x_3\sqrt{F_\pm}}{(r^2\pm1)^2}
-\left(\frac{8c_2x_3^2+c_4x_3\sqrt{F_\pm}}{(r^2\pm1)^2}\cdot H\right),\\
&Q_2=\{D,K_3\pm P_3\}-15x_3-4\left(\left(\frac{c_3x_3}{\sqrt{F_\pm}}+\frac{c_1x_3(r^2\mp1)}{\sqrt{F_\pm}(r^2\pm1)^2}\right)\cdot H\right)\\& -\frac14\left(\frac{c_4(r^2\mp1)((r^2\pm1)^2\mp8x_3^2)}{(r^2\pm1)^2\sqrt{F_\pm}}\cdot H\right)\\&+4\left(\frac{c_7x_3}{\sqrt{F_\pm}}+\frac{c_5x_3(r^2\mp1)}{(r^2\pm1)^2}\right)
+\frac{c_8(r^2\mp1)((r^2\pm1)^2\mp8x_3^2)}{4(r^2\pm1)^2\sqrt{F_\pm}};\end{split}\end{gather}
\begin{gather}\la{last3}\begin{split}
&f=\frac{\tilde r^2\sqrt{F_\pm}}{c_1\sqrt{F_\pm}+c_2(x_3-1)\sqrt{r^2\pm1+2x_3}+c_3(x_3+1)\sqrt{r^2\pm1-2x_3}+c_4(r^2\mp 1)},\\&V=\frac{c_5\sqrt{F_\pm}+c_6(x_3-1)\sqrt{r^2\pm1+2x_3}+c_7(x_3+1)\sqrt{r^2\pm1-2x_3}+c_8(r^2\mp 1)}{c_1\sqrt{F_\pm}+c_2(x_3-1)\sqrt{r^2\pm1+2x_3}+c_3(x_3+1)\sqrt{r^2\pm1-2x_3}+c_4(r^2\mp 1)},\\&
Q_1=\{D,K_3\pm P_3\}-15x_3+\frac{c_6(r^2\pm1)}{\sqrt{r^2\pm1-2x_3}}+\frac{c_7(r^2\pm1)}{\sqrt{r^2\pm1+2x_3}}
+\frac{4c_8}{\sqrt{F\pm}}\\&-\left(\left(\frac{c_2(r^2\pm1)}{\sqrt{r^2\pm1-2x_3}}+
\frac{4c_4}{sqrt{F_\pm}}\right)\cdot H\right)
-\left(\frac{c_3(r^2\pm1)}{\sqrt{r^2\pm1+2x_3}}\cdot H\right) ,\\
&Q_2=\{P_3,K_3\pm P_3\}+\frac{c_6x_3}{\sqrt{r^2\pm1-2x_3}}
+\frac{c_7x_3}{\sqrt{r^2\pm1+2x_3}}+\frac{c_8(r^2\pm 1)}{\sqrt{F_\pm}}\\&-\left(\frac{c_3x_3}{\sqrt{r^2\pm1+2x_3}}\cdot H\right)-\left(\left(\frac{c_2x_3}{\sqrt{r^2\pm1-2x_3}}+\frac{c_4(r^2\pm 1)}{\sqrt{F_\pm}}\right)\cdot H\right);\end{split}\\
\la{last5}\begin{split}&f=\frac{(r^2\pm 1)^2r\tilde r^2}{c_1(r^2\pm 1)^2+c_2r\tilde r^2+c_3x_3(r^2\pm1)^2+c_4\tilde r^2(r^2\mp1)},\\&V=\frac{c_5(r^2\pm 1)^2+c_6r\tilde r^2+c_7x_3(r^2\pm1)^2+c_8\tilde r^2(r^2\mp1)}{c_1(r^2\pm 1)^2+c_2r\tilde r^2+c_3x_3(r^2\pm1)^2+c_4\tilde r^2(r^2\mp1)},\\&Q_1=\{K_3,P_3\}+\frac{c_7x_3(r^2\pm 1)^2-x_3^2(c_6r+c_8(r^2\mp 1))}{r(r^2\pm 1)^2}\\&-\left(\frac{c_3x_3(r^2\pm 1)^2-x_3^2(c_2r+c_4(r^2\mp 1))}{r(r^2\pm 1)^2}\cdot H\right),\\&Q_2=\{D,(K_3\pm P_3)\}-\hat\eta+\frac{2c_6x_3(r^2\mp1)}{(r^2\pm1)^2}+\frac{c_8(r^4\mp r^2+1)-c_7(r^4-1)(r^2\pm1)}{r(r^2\pm1)^2}\\&-\left(\frac{2c_2x_3r(r^2\mp1)-c_3(r^4-1)(r^2\pm1)+c_4(r^4\mp r^2+1)}{r(r^2\pm1)^2}\cdot H\right).\end{split}
\end{gather}

\section{Discussion}
In contrast with the cases of the standard Schr\"odinger equations and 2D Schr\"odinger equations with position dependent mass we still do not have the completed description of second order integrals of motion for 3D PDM Schr\"odinger equations. However, some steps to such description have been already made: the maximally superintegrable sytems and  separable systems have been already classified \cite{Kal5,Kal31},  there are successes in  the classification of nondegenerate and semidegenerate systems \cite{Cap,Mil}. Moreover, the systems invariant w.r.t. three- and two-parametric Lie groups are classified completely \cite{AG,AG1}.

  In the present paper we make the next stem to the complete classification of the mentioned  integrals of motion. Namely, we present all inequivalent quantum mechanical PDM systems which, in addition to the second order integrals of motion, admit the fixed one parameter Lie symmetry group.

As it was shown in \cite{NZ} there are six inequivalent one parametric Lie groups which can be possessed by the PDM systems. We start with one of them,   namely, with the group of rotations around the fixed axis. In other words, we deal with the cylindrically symmetric PDM systems and classify such of them which admit at least one second order integral of motion.

Let us mention that the PDM systems with cylindric symmetry possess a rather extended collection of the mentioned integrals. Namely, we have fixed as much as 66 inequivalent systems and presented their integrals of motion.  In particular we specify 18 superintrgrable and 10 maximally superintgrable systems. They are collected in five tables. In addition, the most cumbersome of them are presented  separately in formulae (\ref{last4})-(\ref{last5}). Notice that any item including the terms "$\pm$" in fact represents two systems one of which corresponds to the sign "$+$" and the other to the sign"$-$".

To optimize calculations we separate the integrals of motion to three qualitatively different subclasses in accordance with their transformation properties with respect to the rotations which by definition leaves  the PDM  Hamiltonians invariant.  The mentioned subclasses include the scalar, vector and tensor versions of the integrals. Moreover, the scalar and tensor integrals of motion can be effective separated in accordance with their parity properties.

The systems admitting one scalar integral of motion are defined up to arbitrary functions which depend on specific variables. Such (integrable) systems are presented in Table 1.

The majority of systems admitting vector or tensor integrals of motion are defined more strictly and includes only arbitrary parameters. The reason of it is that for these subclasses the related PDM system is supposed to admit as minimum two linearly independent integrals of motion. The same is true for the case of superintegrable systems admitting  integrals of motion of arbitrary type.

\begin{center}Table 4.  Inverse   masses, potentials  and  combined integrals of motion for superintegrable systems \end{center}
\begin{tabular}{c c c c}
\hline

\vspace{1.5mm}No&$f$&$V$&\text{Integrals of motion}\\
\hline\\

1\vspace{1 mm}&$\frac{ \tilde rx_3^2}{\tilde rF\left(\frac{r^2\pm1}{x_3}\right)+
c_1x_3^2 (r^2\mp1)F_{\pm }}$&$
\frac{\tilde rG\left(\frac{r^2\pm1}{x_3}\right)+
 {c_2}x_3^2 (r^2\mp1)F_{\pm }} {\tilde rF\left(\frac{r^2\pm1}{x_3}\right)+
c_1x_3^2 (r^2\mp1)F_{\pm }}$&$\begin{array}{c}
\{L_3,(K_\alpha\pm P_\alpha)\}+3\varepsilon_{\alpha\nu}x_\nu\\+2c_1\left(\frac{\varepsilon_{\alpha\nu} x_\nu}{\tilde r}\cdot H\right)-2 {c_2}\frac{\varepsilon_{\alpha\nu} x_\nu}{\tilde r},\\(K_3\pm p_3)^2+\frac{G\left(\frac{r^2\pm 1}{x_3}\right)F_\pm}{x_3^2}\\-\left(\frac{F\left(\frac{r^2\pm 1}{x_3}\right)F_\pm}{x_3^2}\cdot H\right)
\end{array}$\\

\hline

2\vspace{1.5mm}&$\frac{ (r^4-1)^2x_3^2}{(r^2\pm1)^2({c_1}(r^2\mp1)^2+{c_2} x_3^2)-c_3 x_3^2r^2}$&$\frac{ (r^2\pm1)^2({c_4}(r^2\mp1)^2+{c_5} x_3^2)- {c_6} x_3^2r^2}{(r^2\pm1)^2({c_1}(r^2\mp1)^2+{c_2} x_3^2)-c_1 x_3^2r^2}$&$\begin{array}{c}(K_1\pm P_1)^2-(K_2\pm P_2)^2\\+15(x_1^2-x_2^2)\\+c_1\left(\frac{x_1^2-x_2^2}{(r^2\pm1)^2}\cdot H\right)+ {c_2}\frac{x_1^2-x_2^2}{(r^2\pm1)^2},\\\{(K_1\pm P_1),(K_2\pm P_2)\}+15x_1x_2\\+c_1\left(\frac{x_1x_2}{(r^2\pm1)^2}\cdot H\right)+ {c_2}\frac{2x_1x_2}{(r^2\pm1)^2},\\\{L_1,L_2\}-\frac{{c_4} x_1x_2}{x_3^2}+\left(\frac{{c_1} x_1x_2}{x_3^2}\cdot H\right),\\L_1^2
-L_2^2-\frac{c_4 (x_1^2-x_2^2)}{2x_3^2}\\+\left(\frac{c_1( x_1^2-x_2^2)}{2x_3^2}\cdot H\right)\end{array}$\\

\hline

3\vspace{1.5mm}&$\frac1{c_1 \tilde r^2+F(x_3)}$&$\frac{ {c_2}\tilde r^2+G(x_3)}{c_1 \tilde r^2+F(x_3)}$&$\begin{array}{c}P_1P_2-(c_1x_1x_2\cdot H)+ {c_2} x_1x_2,\\P_3^2-\left(F(x_3)\cdot H\right)+G(x_3),\\P_1^2-P_2^2+ {c_2} (x_1^2-x_2^2)-\\(c_1(x_1^2-x_2^2)\cdot H)
\end{array}$\\

\hline

4\vspace{1.5mm}&$\frac{x_3^2}{x_3^2(c_1 r^2+ {c_2})+{c_1}}$&$\frac{x_3^2({c_3} r^2+{c_2})+c_4}{x_3^2(c_1 r^2+ {c_2})+{c_1}}$&$\begin{array}{c}L_1^2-L_2^2-\frac{c_4 (x_1^2-x_2^2)}{2x_3^2}+\left(\frac{{c_1} (x_1^2-x_2^2)}{2x_3^2}\cdot H\right),\\\{L_1,L_2\}-\frac{c_4 x_1x_2}{x_3^2}+\left(\frac{{c_1} x_1x_2}{x_3^2}\cdot H\right)\\P_3^2+{c_3} x_3^2+\frac{c_4}{x_3^2}-\left((c_1 x_3^2+\frac{c_2}{x_3^2})\cdot H\right)\end{array}$\\

\hline

5\vspace{1 mm}&$\frac{\tilde r}{F(x_3)\tilde r+c_1}$&$\frac{G(x_3)\tilde r+ {c_2}}{F(x_3)\tilde r+
c_1}$&$\begin{array}{c}\{L_3,P_\alpha\}-
 {c_2}\frac{\varepsilon_{\alpha\nu}x_\nu}{\tilde r}+c_1\varepsilon_{\alpha\nu}\left(\frac{x_\nu}{\tilde r}\cdot H\right)\\P_3^2-\left(F(x_3)\cdot H\right)+G(x_3)\end{array}$\\

 \hline

6\vspace{1 mm}&$\frac{x_3^2r}{(c_1x_3^2 +c_3)r+2c_2 x_3^2}$&$\frac{(c_5x_3^2 +c_4)r+c_6 x_3^2}{(c_1x_3^2 +c_3)r+2c_2 x_3^2}$&$\begin{array}{c}\{P_\alpha,D\}-2\left(x_\alpha(c_1+\frac{c_2}{r})\cdot H\right)\\+2x_\alpha(2c_5+\frac{c_6}{r}),\\P_3^2-\left(\frac{c_3^2}{x_3^2}\cdot H\right)+\frac{c_4}{x_3^2}\end{array}$\\

\hline

7\vspace{1 mm}&$((r^2\pm1)^2\mp 4x_3^2)^\frac12$&$c_1$&$\begin{array}{c}\{(K_3\pm P_3,P_\alpha\}+P_nx_3x_\alpha P_n,\\\{P_3,(K_3\pm P_3)\}+P_ax_3^2P_a\end{array}$\\
\hline\hline
\end{tabular}

\vspace{2mm}

\newpage

\begin{center}Table 5.  Inverse   masses, potentials  and  integrals of motion for maximally superintegrable systems  \end{center}

\begin{tabular}{c c c c}

\hline
\vspace{1.5mm}No&$f$&$V$&\text{Integrals of motion}\\
\hline\\

1\vspace{1 mm}&$\frac{r^2{\tilde r}}{\tilde rF(r)+c_1  x_3}$&$\frac{\tilde r(G(r)+ {c_2} x_3)}{\tilde rF(r)+c_1  x_3}$&$\begin{array}{c}\{L_3,L_\alpha\}-2\left(\frac{c_1 x_\alpha}{\tilde r}\cdot H\right)+\frac{2 {c_2} x_\alpha}{\tilde r},\\L_1^2+L_2^2-
c_1 (\frac{x_3}{\tilde r}\cdot H)+\frac{ {c_2} x_3}{\tilde r},\\\{K_3,P_\alpha\}+2\varepsilon_{\alpha\nu}DL_\nu\\+\left(\frac{x_\alpha}
{r^2}(x_3F(r)-c_1\tilde r)\cdot H\right)\\-\frac{x_\alpha}{r^2}( G(r)-c_2\tilde r)+P_ax_3x_\alpha P_a\end{array}$\\

\hline

2\vspace{1.5mm}&$\frac{1}{c_1+2c_2x_3+c_3(\tilde r^2+4x_3^2) }$&$\frac{c_4+2c_5x_3+c_6(\tilde r^2+4x_3^2) }{c_1+2c_2x_3+c_3(\tilde r^2+4x_3^2) }$&$\begin{array}{c}P_3^2+2c_5x_3+4c_6x_3^2\\-\left((2c_2x_3+4c_3x_3^2)\cdot H\right),\\\{P_3,D\}-P_nx_3P_n+c_4x_3+\\c_5(2x_3^2+\tilde r^2)+c_6x_3(4x_3^2+3\tilde r^2)\\-\left((c_1x_3+c_2(2x_3^2+\tilde r^2)\right.\\\left.+c_3x_3(4x_3^2+3\tilde r^2))\cdot H\right),\\P_1L_1-P_2L_2\\-(x_1x_2(2c_3 x_3+c_2)\cdot H)\\+x_1x_2(2c_6 x_3+c_5),\\\{P_1,L_2\}+\{P_2,L_1\}\\-((x_1^2-x_2^2)(2c_3 x_3+c_2)\cdot H)\\+(x_1^2-x_2^2)(2c_6 x_3+c_5)\end{array}$\\

\hline

3\vspace{1 mm}&$\frac{\tilde r^2}{c_1+c_2 x_3 \tilde r}$&$\frac{c_3+c_4 x_3 \tilde r}{c_1+c_2 x_3 \tilde r}$&$\begin{array}{c}\{L_3,P_\alpha\}+c_1\varepsilon_{\alpha\nu}\left(\frac{x_\nu}{\tilde r}\cdot H\right)-
 {c_3}\frac{x_2}{\tilde r},\\P_3^2-\left(c_2x_3\cdot H\right)+c_4x_3,\\\{P_3,D\}+\left(\left(\frac{c_1}{x_3^2}-c_2(x_3^2+\frac{\tilde r^2}2\right)\cdot H\right)\\-\frac{c_3}{x_3^2}-c_4(x_3^2+\frac{\tilde r^2}2)\end{array}$\\

 \hline

4\vspace{1 mm}&$\frac{1}{c_1r^2+2c_2x_3+c_3}$&$\frac{c_4r^2+2c_5x_3+c_6}{c_1r^2+2c_2x_3+c_3}$&
$\begin{array}{c}P_3P_\alpha-\left(x_\alpha(c_1x_3+c_2)\cdot H\right)\\+x_\alpha(c_4x_3+c_5),\\
P_3^2-\left((c_1x_3^2+2c_2x_3)\cdot H\right)\\+c_4x_3^2+2c_5x_3,\\
P_1P_2-c_1(x_1x_2\cdot H)+c_4x_1x_2,\\(P_1^2-P_2^2)+c_4(x_1^2-x_2^2)\\-c_1((x_1^2-x_2^2)\cdot H)\\\end{array}
$\\

\hline

5\vspace{1.5mm}&$\frac{x_3^2}{c_1 x_3^2 r^2+{c_2}}$&$\frac{{c_3} x_3^2r^2+{c_4}}{c_1 x_3^2 r^2+{c_2}}$&$\begin{array}{c}P_1P_2-c_1(x_1x_2\cdot H)+{c_3} x_1x_2,\\P_1^2-P_2^2+{c_3} (x_1^2-x_2^2)-{c_1}((x_1^2-x_2^2)\cdot H),\\\{L_1,L_2\}+\left(\frac{2{c_2} x_1x_2}{x_3^2}\cdot H\right)-\frac{2{c_4} x_1x_2}{x_3^2},\\L_1^2-L_2^2+\left(\frac{{c_2} (x_1^2-x_2^2)}{x_3^2}\cdot H\right)-\frac{{c_4} (x_1^2-x_2^2)}{x_3^2},\\P_3^2-\left((c_1 x_3^2+\frac{c_2}{x_3^2})\cdot H\right)+{c_3} x_3^2+\frac{c_4}{x_3^2}
\end{array}$\\

\hline\hline
\end{tabular}

\vspace{1mm}

\newpage

\begin{center}Table 5 (continued).  Inverse   masses, potentials  and  integrals of motion for maximally superintegrable systems  \end{center}

\begin{tabular}{c c c c}

\hline
\vspace{1.5mm}No&$f$&$V$&\text{Integrals of motion}\\
\hline\\

6\vspace{1.5mm}&$\frac{\tilde r^2\sqrt{F_\pm}}{c_1\sqrt{F_\pm}+c_2(r^2\mp 1)}$&$\frac{c_3\sqrt{F_\pm}+c_4(r^2\mp1)}{c_1\sqrt{F_\pm}+c_2(r^2\mp1)}$&$\begin{array}{l}K_3^2- P_3^2-\hat\eta-\frac{2c_4(r^2\mp1)}{\sqrt{F_\pm}}
+\left(\frac{2c_2(r^2\mp1)}{\sqrt{F_\pm}}\cdot H\right),\\
\{D,(K_3\pm P_3)\}-15x_3\mp\frac{4c_4x_3}{\sqrt{F_\pm}}
\pm\left(\frac{4c_2x_3}{\sqrt{F_\pm}}\cdot H\right),\\(K_3\pm P_3)^2-\hat\eta\end{array}$\\

\hline

7\vspace{1 mm}&$\frac{r^2\tilde r^2}{c_1{r^2}+c_2r^4\tilde r^2+c_3r^2\tilde r^2}$&$\frac{c_4{r^2}+c_5r^4\tilde r^2+c_6r^2\tilde r^2}{c_1{r^2}+c_2r^4\tilde r^2+c_3r^2\tilde r^2}$&$\begin{array}{c}
\{P_3,K_3\}-2P_ax_3P_a\\+\left((\frac{c_2x_3^2}{\tilde r^2}-c_3 x_3^2)\cdot H\right)\\-\frac{c_5x_3^2}{\tilde r^2}+c_6 x_3^2,\\L_1^2+L_2^2-
(c_1\frac{r^2}{\tilde r^2}\cdot H)+c_4\frac{r^2}{\tilde r^2},\\P_3^2-\left(c_2x_3^2\cdot H\right)+c_5x_3^2\end{array}$\\

\hline

8\vspace{1.5mm}&$\frac1{ {c_2}(\tilde r^2+4x_3^2)+2{c_3} x_3+c_1}$&$\frac{{c_6} x_3+{c_5}(\tilde r^2+4x_3^2)+c_4}{ {c_2}(\tilde r^2+4x_3^2)+2{c_3} x_3+c_1}$&$\begin{array}{c}P_1L_1-P_2L_2+x_1x_2(2{c_5} x_3+{c_6})\\-(x_1x_2(2 {c_2} x_3+{c_3})\cdot H),\\\{P_1,L_2\}+\{P_2,L_1\}+(x_1^2-x_2^2)(2{c_5} x_3+{c_6})\\-((x_1^2-x_2^2)(2 {c_2} x_3+{c_3})\cdot H)\\P_3^2-2\left((2 {c_2} x_3^2+{c_3} x_3)\cdot H\right)\\+4{c_5} x_3^2+{c_6} x_3,\\P_1P_2- {c_2} (x_1x_2\cdot H)+{c_5} x_1x_2,\\P_1^2-P_2^2-c_1((x_1^2-x_2^2)\cdot H)+c_4 (x_1^2-x_2^2)
\end{array}$\\

\hline

9\vspace{1 mm}&$\frac{x_3^2}{c_1x_3^2(x_3^2+4\tilde r^2)+c_2+c_3x_3^2}$& $\frac{c_4+c_5x_3^2(x_3^2+4\tilde r^2)+c_6x_3^2}{c_1x_3^2(x_3^2+4\tilde r^2)+c_2+c_3x_3^2}$&$\begin{array}{c}\{P_3,L_\alpha\}+
2\frac{\varepsilon_{\alpha\nu}x_\nu(c_4-c_5x_3^4)}{x_3^2}+2\left(\frac{\varepsilon_{\alpha\nu}x\nu(c_1x_3^4-c_2)}{x_3^2}\cdot H\right),\\P_3^2-\left((\frac{c_1}{x_3^2}+\frac{c_2}{x_3^2})\cdot H\right)\
+\frac{c_4}{x_3^2}+\frac{c_5}{x_3^2},\\ P_1P_2-c_1(x_1x_2\cdot H)+c_4 x_1x_2,\\P_1^2-P_2^2-c_1((x_1^2-x_2^2)\cdot H)+c_4 (x_1^2-x_2^2)\end{array}$\\

\hline

10\vspace{1 mm}&$\frac{x_3^2(r^2\pm1)}{c_1(r^2\pm1)
+c_2x_3^2 }$&$ \frac{c_3(r^2\pm1)
+c_4x_3^2}{c_1(r^2\pm1)
 +c_2x_3^2}$&$\begin{array}{c}\{D,(K_\alpha\pm P_\alpha)\}-15x_\alpha-\frac{2x_\alpha c_4(r^2\mp1)}{(r^2\pm1)^2}\\
+\left(\frac{2x_\alpha c_1((r^2\mp1)}{(r^2\pm1)^2}\cdot H\right),\\\{(K_3\pm P_3),L_\alpha\}+3\varepsilon_{\alpha\nu}x_\nu
-\frac{2c_4\varepsilon_{\alpha\nu}x_\nu(r^2\pm1)}{x_3^2}\\+2c_1\left(\frac{\varepsilon_{\alpha\nu}x_\nu(r^2\pm1)}{x_3^2}\cdot H\right),
\\\{L_1,L_2\}+\left(\frac{2c_1 x_1x_2}{x_3^2}\cdot H\right)-\frac{2c_3 x_1x_2}{x_3^2},\\L_1^2-L_2^2
+\left(\frac{2c_1 (x_1^2-x_2^2)}{x_3^2}\cdot H\right)-\frac{2c_3  (x_1^2-x_2^2)}{x_3^2}

\end{array}$\\

\hline\hline
\end{tabular}
\vspace{3mm}

Thus we have made an essential step to the complete classification of the 3D PDM systems admitting second order integrals of motion. In spite of that we consider only one out of six inequivalent one parametric Lie groups which can be accepted by such systems, this step is very important since the number of found systems is very large, maybe more large then the total number of all systems admitting the other inequivalent Lie symmetries. The latter statement is supported by our computing experiments.

We believe that the presented  classification is complete. However, the determining equations which we solve to find the inequivalent systems and their integrals of motion are rather complicated systems of partial differential equations with variable coefficients, and there is a danger to overlook some special solutions additional to the found generic ones. That is why  we present the mentioned determining equations explicitly. And the absolutely rigorous statement (which, however, is tot constructive)  is that the discussed systems should include the arbitrary elements which are functions solving these equations.

The next planned steps to the complete classification of the integrals of motion admitted by the 3D PDM systems presuppose the classifications of the systems which possess symmetries with respect to the remaining inequivalent one parametric Lie groups for such systems specified in \cite{NZ}. Finally, we plane to classify such integrals of motion for the systems which have no Lie symmetry. The latter problem appears not to be catastrophically complicated thanks to the existence of rather strong equivalence group.

{\bf Acknowledgement.} I am indebted with Universit\'a del Piemonte Orientale and and Dipartimento di Scienze e Innovazione Tecnologica for the extended stay as Visiting Professor.

\end{document}